\documentclass[10pt, twocolumn]{IEEEtran}
\usepackage{amssymb}
\usepackage{amsthm}
\usepackage{amsmath}
\usepackage{cite}
\usepackage{amsfonts}
\usepackage{float}
\usepackage{graphicx}
\usepackage{color}
\usepackage{mathtools}
\usepackage{xcolor}
\usepackage{etoolbox}
\usepackage{ifthen}
\usepackage{bm}
\usepackage{algorithm}
\usepackage{algpseudocode}
\usepackage{nccmath}
\usepackage{enumitem}
\usepackage{framed}
\usepackage{threeparttable}
\usepackage{booktabs}
\usepackage{multirow}
\usepackage{subfigure}
\usepackage{makecell}
\usepackage{widetext}
\usepackage{bbm}
\usepackage[caption=false]{subfig}

 \theoremstyle{definition}

\DeclareMathAlphabet\mathbfcal{OMS}{cmsy}{b}{n}

\newcommand{\todo}[1]{\iftrue \textcolor{red}{#1} \fi}
\newcommand{\Figref}[2]{\mbox{Fig. \ref{#1}#2}}

\newcommand{\p}[0]{\mathrm{p}}
\newcommand{\q}[0]{\mathrm{q}}

\graphicspath{ {./images/} }

\begin{document}
\title{Bayesian Deep End-to-End Learning for MIMO-OFDM System with Delay-Domain \\ Sparse Precoder} \author{ Nilesh~Kumar~Jha, \IEEEmembership{Graduate Student member,~IEEE,} Huayan~Guo,\IEEEmembership{ member,~IEEE,} and~Vincent~K.~N.~Lau,~\IEEEmembership{Fellow,~IEEE}%
\normalsize \thanks{This work is supported by in part by the  Research Grants Council of Hong Kong  under Project 16213119,
in part by the Research Grants Council under the Areas of Excellence scheme grant AoE/E-601/22-R and AoE/E-101/23-N
in part by the National Natural Science Foundation of China under Grant 62101472. (Corresponding author: Vincent K. N. Lau)}%
\thanks{Nilesh Kumar Jha, Huayan Guo, and Vincent K. N. Lau are with the Department of Electronic and Computer Engineering, The Hong Kong University of Science and Technology, Clear Water Bay, Kowloon, Hong Kong (e-mail: { nkjha@connect.ust.hk;  eeguohuayan@ust.hk; eeknlau@ece.ust.hk}).}%

}

\maketitle
\IEEEpeerreviewmaketitle
\begin{abstract}
This paper introduces a novel precoder design aimed at reducing pilot overhead for effective channel estimation in multiple-input multiple-output orthogonal frequency division multiplexing (MIMO-OFDM) applications utilizing high-order modulation. We propose an innovative demodulation reference signal scheme that achieves up to an 8x reduction in overhead by implementing a delay-domain sparsity constraint on the precoder. Furthermore, we present a deep neural network (DNN)-based end-to-end architecture that integrates a propagation channel estimation module, a precoder design module, and an effective channel estimation module. Additionally, we propose a Bayesian model-assisted training framework that incorporates domain knowledge, resulting in an interpretable datapath design. Simulation results demonstrate that our proposed solution significantly outperforms various baseline schemes while exhibiting substantially lower computational complexity.
\end{abstract}


\section{Introduction}

The study of multiple-input multiple-output (MIMO) communication combined with orthogonal frequency division multiplexing (OFDM) has a long-standing significance in wireless communication, particularly in WiFi and millimeter-wave backhaul \cite{heath2016overview, bolcskei2006mimo, harkat2022survey}. To address the high data rate requirements of emerging applications such as extended reality (XR) and virtual reality (VR), the forthcoming WiFi protocol 802.11bn aims for a throughput of up to 100 Gbps by increasing the modulation order to 4096 and expanding the number of spatial data streams. However, effective channel recovery necessitates that the number of pilot symbols matches the data streams, while the payload symbol length decreases with a fixed frame bit length as the modulation order increases. These conditions significantly undermine spectral efficiency, making pilot reduction essential. Additionally, demodulating 4K-QAM requires a high signal-to-interference-plus-noise ratio (SINR), which demands high-quality channel recovery for effective decorrelator design to mitigate interference among multiple streams. Therefore, developing techniques that ensure accurate channel state information at the receiver (CSIR) with reduced pilot overhead and limited transmission power is crucial, yet remains inadequately addressed in current research.


Improving channel state information at the transmitter (CSIT) with reduced pilot overhead has been achieved through compressed sensing, leading to sparse channel recovery algorithms \cite{liu2016exploiting, liu2016joint, liu2018downlink} that utilize angular sparsity in narrowband MIMO channels. In MIMO-OFDM systems, wideband operation reduces pilot overhead by leveraging delay-domain sparsity, where the delay spread of the channel impulse response is smaller than the OFDM length. For example, 5G-NR transmits multiple orthogonal pilots within a single OFDM symbol across uniformly sampled sub-carriers \cite{dahlman20205g}, facilitating least squares estimation. Non-uniformly sampled pilot schemes \cite{mohammadian2016deterministic, li2022adaptive, qi2011optimized, chen2013efficient} further exploit delay-domain sparsity with sparse channel recovery algorithms. Recent studies \cite{wang2022channel, srivastava2021bayesian, zhu2019channel, chen2018massive} have enhanced pilot reduction by combining angular and delay-domain sparsity in MIMO-OFDM channel estimation. However, these approaches cannot address CSIR, which represents the effective channel formed by the combination of the propagation channel and the transmit precoder. The effective channel lacks sparsity in both domains due to traditional MIMO-OFDM precoder designs: the precoded MIMO data streams do not transmit in the null space of the angular domain, and the precoder does not exhibit sparsity in the delay domain.

Most traditional MIMO precoding methods, as noted in \cite{christensen2008weighted, shi2011iteratively, perahia2013next, ali2017beamforming, tiwari2023advancing, ravindran2008limited, sung2009generalized, joham2005linear, moon2014channel, zhang2009robust, vucic2009robust, dabbagh2008multiple, fritzsche2013robust}, design precoders independently for each subcarrier, assuming that the receiver operates perfectly. For instance, the weighted minimum mean squared error (WMMSE) algorithm \cite{christensen2008weighted, shi2011iteratively} maximizes the weighted sum rate per subcarrier, while other methods minimize the error vector magnitude (EVM) \cite{joham2005linear} or use singular value decomposition (SVD) with transmit power minimization \cite{perahia2013next, ali2017beamforming, tiwari2023advancing}. These subcarrier-independent designs lead to a non-sparse effective channel, making existing pilot reduction schemes ineffective. Recent studies \cite{precodedelay2019WCL, zhang2023cross} design precoders in the spatial-delay domain to maximize overall sum-rate, primarily to reduce optimization complexity. While this approach can naturally yield a sparse delay-domain effective channel, its significance for pilot reduction has not been acknowledged. Furthermore, these designs often result in suboptimal performance due to varying achievable rates among subcarriers, as all subcarriers assigned to the same receiver typically use the same modulation and coding scheme (MCS). Therefore, a novel precoding design that enhances end-to-end performance while preserving delay-domain sparsity is highly desirable.



Recently, deep neural network (DNN)-based end-to-end MIMO CSIT feedback and precoding design has emerged as a promising approach for software-defined native 6G networks. Most existing research focuses on joint MIMO CSIT feedback and precoding design under the assumption of perfect CSIR, with the aim of addressing imperfect CSIT and reducing computational complexity. Early attempts \cite{sohrabi2021deep, jang2019deep, wu2022deep, zhang2022deep, attiah2022deep, pellaco2020deep} have optimized black-box DNN precoding functions to maximize the average received signal-to-noise ratio (SNR) or achievable sum rate with compressed CSIT. To enhance performance, model-driven DNN precoders have been developed by unrolling iterative algorithms from traditional MIMO designs for sum-rate maximization \cite{jang2022deep, shi2023robust, jin2023model}. However, recognizing that sum-rate is not an ideal metric, a hybrid precoding DNN has been optimized in \cite{hu2021two} using the final demodulation bit error rate (BER) as the training criterion. Nonetheless, the assumption of perfect CSIR leads to designs that neglect the necessity for sparse precoding, which is critical for effective CSIR estimation. Furthermore, the loss functions used in most studies primarily focus on the performance of the final output, leaving the quality of intermediate variables in the data path uncertain. This lack of interpretability poses additional challenges for practical deployment.

\begin{figure}
    \centering
	    \subfigure[TDD transmission frames in a timeline]    {
	    	\includegraphics[trim={0 0 0 0},clip,width=0.95\linewidth]{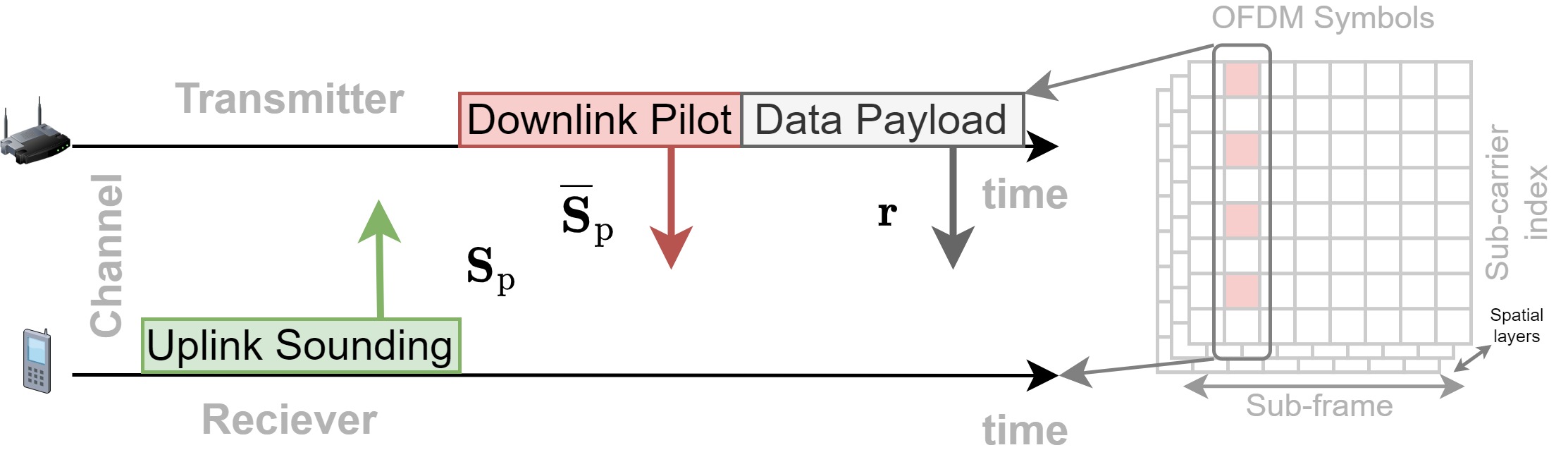}
	    	}
	    \subfigure[Key signal processing modules in the TX and RX]    {
	    	\includegraphics[trim={0 0 0 0},clip,width=0.95\linewidth]
{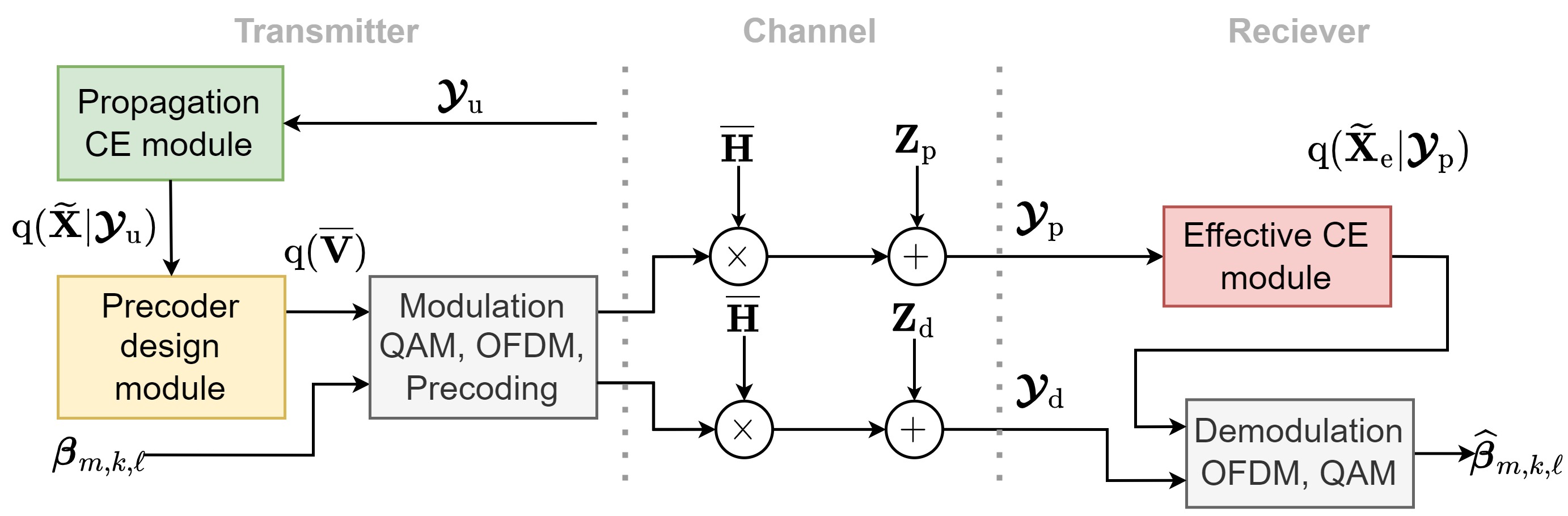}
		}
    \caption{Illustration of the time division duplexing (TDD) transmission scheme and the key signal processing modules. 
    }\label{fig::signal-model}
\end{figure}

This paper introduces a model-assisted end-to-end DNN solution for a time-division duplexing (TDD) MIMO-OFDM system, as shown in \Figref{fig::signal-model}. By utilizing implicit feedback for CSIT, we simplify the feedback process, enabling us to concentrate on the pilot reduction of CSIR estimation. Our solution comprises three main modules: the propagation channel estimation module, the precoder design module, and the effective channel estimation module. We also formulate a variational Bayesian inference (VBI) problem to train the weights of the overall DNN model in a model-assisted manner, ensuring the interpretability of key variables in the data path. The key contributions of this work are summarized as follows:
\begin{enumerate}
\item \textbf{DNN Solution with Delay-Domain Sparse Precoder Design for Pilot Reduction}: We propose a novel DNN-based MIMO-OFDM precoder design that achieves effective channel estimation with up to 8x reduction in pilot overhead by enforcing sparsity in the delay domain. The precoding design is formulated as an average received EVM minimization problem, with a constraint limiting the precoder to a few delay taps. We then introduce an iterative algorithm using block coordinate ascent (BCD). Finally, we apply the DNN unrolling technique \cite{monga2021algorithm} to create a model-assisted DNN precoder based on this iterative algorithm. This approach yields a sparse effective channel with minimal performance loss due to the sparse constraint on the precoder design.

\item \textbf{End-to-End MIMO-OFDM DNN Transceiver Design for Lower BER with Reduced Pilot Overhead}:
In addition to the precoder design module, we propose an end-to-end design for the MIMO-OFDM transceiver. This design incorporates a propagation channel estimation module at the transmitter to estimate CSIT for precoder design by leveraging angular-delay domain sparsity. Simultaneously, we develop an effective channel estimation module at the receiver to support MIMO decorrelator design and 4K-QAM demodulation. These three DNN modules are jointly trained in an end-to-end manner, focusing on minimizing the hard-decision BER. Unlike traditional approaches that optimize intermediate criteria, such as EVM, under the assumption of perfect CSIT and CSIR, our end-to-end training implicitly addresses the challenges of imperfect CSIT and CSIR in precoder design while ensuring that the BER metric aligns directly with the overall design objective.

\item \textbf{Variational Bayesian Training Framework for an Interpretable DNN Datapath}:
To address the issue of non-interpretability, we propose a variational Bayesian training framework that systematically designs Bayesian models for all key variables in the DNN datapath. Our framework integrates a generative model with likelihood and prior components, incorporating wireless domain knowledge into the DNN functions. We utilize Markov prior models for delay-domain CSIT and CSIR to capture clustered sparsity and apply a cross-entropy distortion likelihood to evaluate the BER. In addition, inference models connect the outputs of each DNN module to the generative model, enabling the integration of all trainable parameters into the loss function by minimizing the Kullback-Leibler (KL) divergence between the generative and inference models. This approach enhances DNN training by ensuring that all key variables accurately represent their intended physical meanings, ultimately improving the effectiveness of the end-to-end DNN solution.

\end{enumerate}

\section{MIMO-OFDM System Model}
We consider a MIMO-OFDM system operating in TDD mode across $K$ subcarriers, where the transmitter (TX) and receiver (RX) have $N_\mathrm{t}$ and $N_\mathrm{r}$ antennas, respectively, supporting up to $L \leq \mathrm{min}(N_\mathrm{t}, N_\mathrm{r})$ parallel spatial streams. In this work, we use the ray-tracing channel model \cite{forenza2007simplified} to represent a wireless propagation scenario with limited scattering. The channel matrix $\mathbf{H}_k \in \mathbf{C}^{N_\mathrm{r} \times N_\mathrm{t}}$ for the $k$-th sub-carrier is given as follows:
\begin{equation}
\mathbf{H}_k = \sqrt{\frac{N_\mathrm{r}N_\mathrm{t}}{N_\mathrm{c}N_\mathrm{ray}}} \sum_{c=1}^{N_\mathrm{c}}\sum_{j=1}^{N_{\mathrm{ray},c}} \alpha_{cj} \mathbf{a}_\mathrm{r}(\theta_{cj}^\mathrm{r})\mathbf{a}_\mathrm{t}^\mathrm{H}(\theta_{cj}^\mathrm{t}),
\end{equation}
where  $N_\mathrm{c}$ is the number of clusters,  $N_{\mathrm{ray},c}$ is the number of rays within the $c$-th cluster, and $\{\alpha_{cj}, \theta_{cj}^\mathrm{r}, \theta_{cj}^\mathrm{t} \}$ are the complex gain, angle of arrival (AoA) and angle of departure (AoD) for the $j$-th ray in $c$-th cluster.

\begin{figure}
    \centering
	    	\includegraphics[trim={0 0 0 0},clip,width=0.8\linewidth]{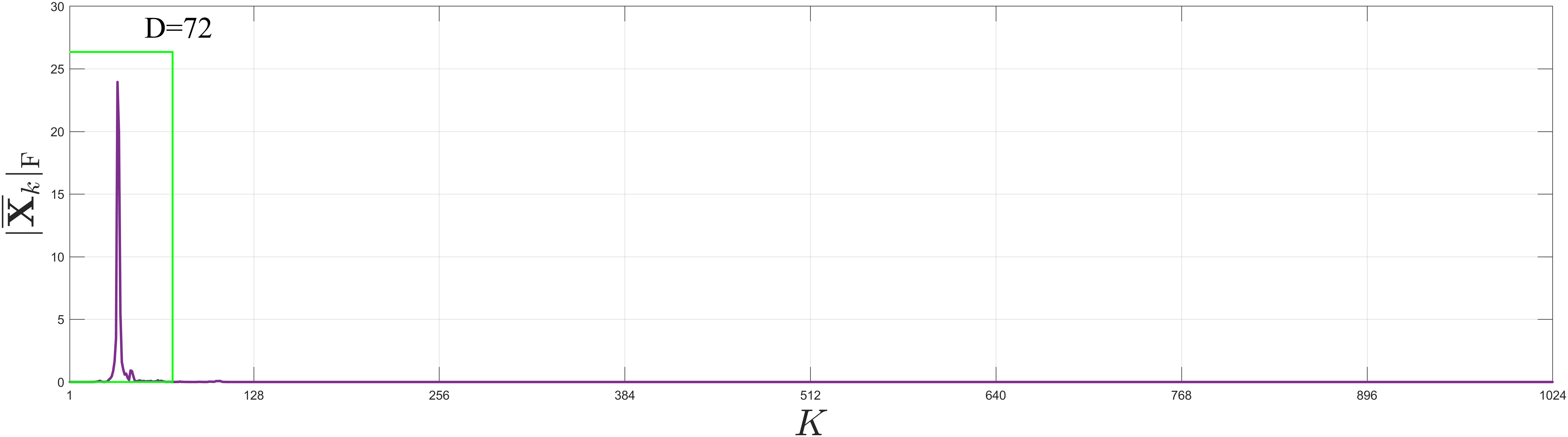}
    \caption{Illustration of delay domain sparsity with limited maximum delay spread ( $D=72$, $K=1024$).}\label{fig::sparsity}
\end{figure}

Denoting the stacked frequency-domain channel by $\mathbf{\overline{H}} = [\mathbf{H}_1^{\rm T}, \cdots, \mathbf{H}_K^{\rm T}]^{\rm T} \in \mathbb{C}^{K N_{\rm r} \times N_{\rm t}}$, the stacked delay-domain channel can then be expressed by:
\begin{equation}\label{eq::feed_start}
\mathbf{\overline{X}} = (\mathbf{F}^\mathrm{H} \otimes \mathbf{I}_{N_\mathrm{r}}) \mathbf{\overline{H}},
\end{equation}
where $\mathbf{F} \in \mathbb{C}^{K \times K}$ is the $K$ point DFT matrix. It is noted that as the various rays arrive within a limited time, the channel exhibits limited maximum delay spread, denoted by $D$, as illustrated in \Figref{fig::sparsity}{}. Consequently, only the first $D N_\mathrm{r}$ rows of $\mathbf{\overline{X}}$ contain non-zero values, which is referred to as delay domain sparsity. The truncated non-zero sub-matrix of $\mathbf{\overline{X}}$ can be expressed as follows:
\begin{equation}\label{eq::feed_start_trunc}
\mathbf{\widetilde{X}} = (\widetilde{\mathbf{F}}_D^\mathrm{H}  \otimes \mathbf{I}_{N_\mathrm{r}})\mathbf{\overline{H}},
\end{equation}
where $\widetilde{\mathbf{F}}_D \in \mathbb{C}^{K \times D}$ is the partial DFT matrix,  created by truncating $\mathbf F$ to keep only the first $D$ columns.

The overall TDD transmission frames are shown in \Figref{fig::signal-model}(a). The RX first sends uplink sounding reference signals (SRS) for CSIT estimation. Using this estimation, the TX designs the precoding matrix, taking advantage of TDD channel reciprocity for downlink data transmission. Additionally, demodulation reference signals (DMRS) are sent along with the data payloads to facilitate effective channel estimation at the receiver. The rest of this section will outline the signal models and processing modules involved in the transmission procedure described above.

\subsection{{Uplink Sounding for Implicit CSIT Feedback}}\label{sec:SRS_signal}
During uplink channel sounding, the RX transmits SRS over $M$ OFDM symbols.  The received pilot  at the TX for the $k$-th sub-carrrier and $m$-th OFDM symbol is given by:
\begin{equation}\label{eq::cs0}
\begin{aligned}[b]
\mathbf{y}_{\mathrm{u},  m,k} &= \mathbf{H}_{k}^\mathrm{T} \mathbf{s}_{\mathrm{u}, m,k } + \mathbf{z}_{\mathrm{u},m, k},
\end{aligned}
\end{equation}
where $\mathbf{s}_{\mathrm{u},m,k} = [s_{\mathrm{u}, m,k,1}, \cdots, s_{\mathrm{u},  m,k, N_\mathrm{r}}]^\mathrm{T} \in \mathbb{C}^{N_\mathrm{r}}$ is the pilot at each subcarrier, $\mathbf{z}_{\mathrm{u}, k}$ is the additive white Gaussian noise (AWGN) with $\mathbf{z}_{\mathrm{u}, m,k} \sim \mathcal{CN}(0, \sigma_\mathrm{u}^2 \mathbf{I}_{N_\mathrm{t}})$.

\begin{figure}[t]
    \centering
    \subfigure[Uniformly sampled FDM pilot scheme]    {
	    	\includegraphics[trim={0 0 0 0},clip,width=0.8\linewidth]{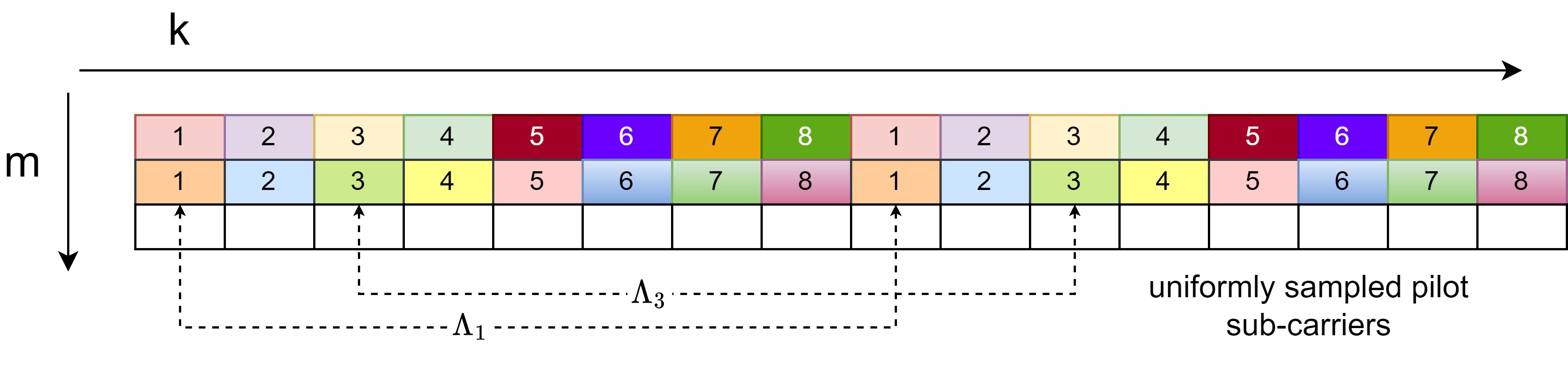}
	    	}
	    \subfigure[Delay domain representation for one pilot stream]    {
	    	\includegraphics[trim={0 0 0 0},clip,width=0.8\linewidth]{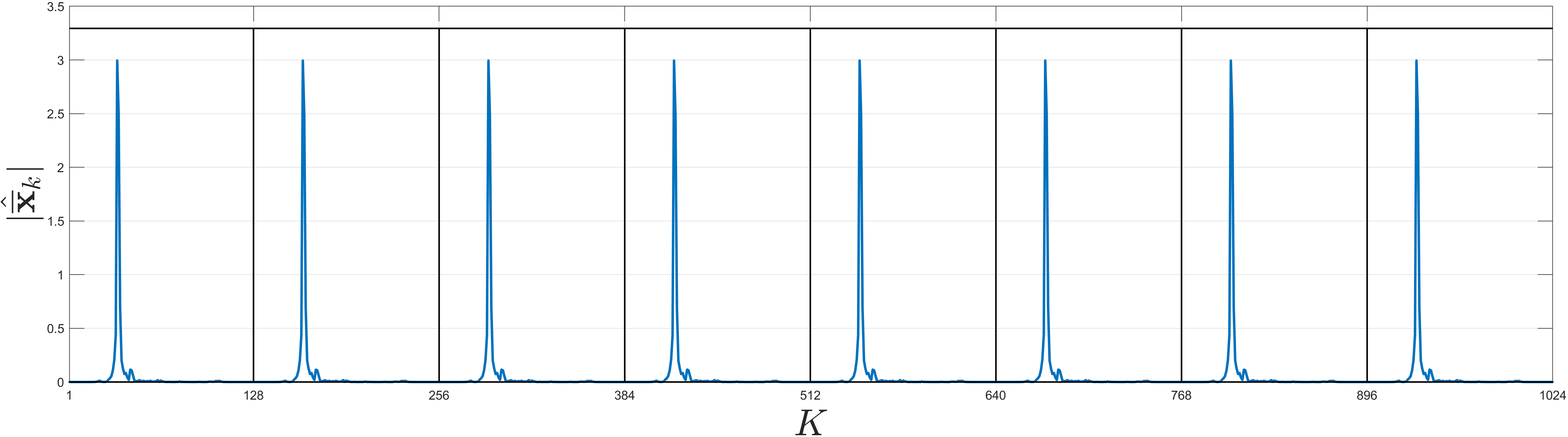}
		}
    \caption{Pilot allocation and resulting delay domain characteristics for $A_\mathrm{u}=8$ and $K=1024$. }\label{fig::pilot-allocation}
\end{figure}

\subsubsection{{FDM Pilot Scheme for SISO Channel Estimation by Exploiting Delay-Domain Sparsity}}
For simplicity, we start with the SISO channel estimation as a toy example. The $m$-th received OFDM symbol is given by:
\begin{equation}\label{eq::SISO_y}
\begin{aligned}[b]
\overline{\bf y}_{\mathrm{u},m} &=  {\rm {diag}}( \overline{\bf s}_{\mathrm{u},m}) \overline{\bf h} + \overline{\bf z}_{\mathrm{u},m},
\end{aligned}
\end{equation}
where $\overline{\bf h} \in \mathbb{C}^K$ is the stacked frequency-domain channel, $\overline{\bf z}_{\mathrm{u},m} \sim \mathcal{CN}(0, \sigma_\mathrm{u}^2 {\bf I}_K)$ is the AWGN.
A uniformly sampled frequency division multiplexing (FDM) pilot scheme is proposed in 5G-NR \cite{dahlman20205g}. This scheme supports multiple orthogonal pilot streams, denoted by $A_\mathrm{u}$, within a single OFDM symbol, as illustrated in \Figref{fig::pilot-allocation}(a). The set of sub-carrier indices allocated for the $a$-th pilot stream is given by:
\begin{equation}
{ \Omega}_a = \{k | \, (k-a) \% A_\mathrm{u} = 0, 1\leq k \leq K\}.
\end{equation}
Then, denote the $a$-th pilot sequence by $\overline{\bf s}_a \in \mathbb{C}^K$, whose $k$-th element is given by
\begin{equation}
\overline{ s}_{a,k} = \begin{cases}
				1,& {\rm{if}} \;\; k \in { \Omega}_a, \\
				0,&  \mathrm{otherwise}.
			\end{cases}
\end{equation}
For SISO channel estimation, only one pilot is needed, so we have $\overline{\bf s}_{\mathrm{u},1}=\overline{\bf s}_a$, where $a$ can be any value selected from the set $ \{1,2,\cdots,A_\mathrm{u}\}$. The received signal $\overline{\bf y}_{\mathrm{u},1}$ becomes
\begin{equation}\label{eq::SISO_ys}
\begin{aligned}[b]
\overline{\bf y}_{\mathrm{u},1} &=  {\rm {diag}}( \overline{\bf s}_a ) \overline{\bf h} + \overline{\bf z}_{\mathrm{u},1}.
\end{aligned}
\end{equation}
One may simply project $\overline{\bf y}_{\mathrm{u},1}$ into delay domain to obtain:
\begin{equation}\label{eq::SISO_y_project}
\begin{aligned}[b]
\hat{\overline{\bf x}} &=  (\mathbf{F}_a^\mathrm{H} \mathbf{F}_a)^{-1}\mathbf{F}^\mathrm{H} \overline{\bf y}_{\mathrm{u},1},
\end{aligned}
\end{equation}
where $\mathbf{F}_a \in \mathbb{C}^{\frac{K}{A_\mathrm{u}} \times K} $ is the partial DFT matrix selecting the $\frac{K}{A_\mathrm{u}}$ rows of $\bf F$ according to ${ \Omega}_a$.
{\Figref{fig::pilot-allocation}{}(b)} illustrates the amplitudes of $\hat{\overline{\bf x}}$. It shows that the uniform pilot scheme introduces $A_\mathrm{u}-1$ sampling artifacts in the delay domain. Since the delay-domain channel is sparse, occupying the first $D$ taps as depicted in \Figref{fig::sparsity}, one can recover $\overline{\bf h}$ using an anti-aliasing filter in the delay domain if $K\geq DA$. This is expressed as:
\begin{equation}\label{eq::est_h_siso}
\hat{\overline{\bf h}} = {\widetilde{\bf F}_D} ({\widetilde{\bf F}}_{D,a}^\mathrm{H} {\widetilde{\bf F}}_{D,a})^{-1} {\widetilde{\bf F}}_D^\mathrm{H} \overline{\bf y}_{\mathrm{u},1},
\end{equation}
where ${\widetilde{\bf F}}_{D,a} \in \mathbb{C}^{\frac{K}{A_\mathrm{u}} \times D} $ selects the $\frac{K}{A_\mathrm{u}}$ rows of ${\widetilde{\bf F}}_D$ according to ${ \Omega}_a$.
Therefore, we can conclude that the uniform FDM pilot schemes can support orthogonal pilot streams up to $A_\mathrm{u} = \lfloor K/D \rfloor$.

\subsubsection{{FDM Pilot Scheme for SIMO Channel}}
The SISO channel estimation can be easily extended to the SIMO cases, where the received signal in \eqref{eq::SISO_ys} becomes
\begin{equation}\label{eq::SIMO_ys}
\begin{aligned}[b]
\overline{\bf Y}_{\mathrm{u},1} &=  {\rm {diag}}( \overline{\bf s}_a ) \overline{\bf H} + \overline{\bf Z}_{\mathrm{u},1},
\end{aligned}
\end{equation}
where $\mathbf{\overline{Y}}_{\mathrm{u}, 1} \in \mathbb{C}^{K \times N_\mathrm{t}}$ and  $\overline{\bf H} \in \mathbb{C}^{K \times N_\mathrm{t}}$. The channel can be estimated by utilizing the same filter shown in \eqref{eq::est_h_siso}:
\begin{equation}\label{eq::est_h_simo}
\hat{\overline{\bf H}} = {\widetilde{\bf F}}_D ({\widetilde{\bf F}}_{D,a}^\mathrm{H} {\widetilde{\bf F}}_{D,a})^{-1} {\widetilde{\bf F}}_D^\mathrm{H} \overline{\bf Y}_{\mathrm{u},1}.
\end{equation}


%

\subsubsection{{FDM Pilot Scheme for MIMO Channel}}\label{sec::mimo_fdm}
We finally consider the MIMO channel estimation when $N_\mathrm{r}>1$, which costs $M = \lceil \frac{N_\mathrm{r}}{A_\mathrm{u}}\rceil$ OFDM symbols.
To address the issue of multiple transmit antennas, we consider a unitary matrix ${\bf S}_{\rm p}=[{\bf s}_{{\rm p},1},{\bf s}_{{\rm p},2},\cdots,{\bf s}_{{\rm p},N_\mathrm{r}}]^\mathrm{T} \in \mathbb{C}^{N_\mathrm{r} \times N_\mathrm{r}}$. We can map ${\bf s}_{{\rm p},n_\mathrm{r}}$ to the  $N_\mathrm{r}$ orthogonal pilot sequences using the following indicator:
\begin{equation}
\alpha_{m, k, n_\mathrm{r}} = \begin{cases}
				1,& {\rm{if}}\; \lceil \frac{n_\mathrm{r}}{A_\mathrm{u}}\rceil=m \; {\rm{and}} \,\,\, k \in \Omega_{n_\mathrm{r}\%A_\mathrm{u}},\\
				0,&  \mathrm{otherwise}.
			\end{cases}
\end{equation}
Then, letting $\mathbf{s}_{\mathrm{u}, m, k }=\sum_r \alpha_{m, k, n_\mathrm{r}} {\bf s}_{{\rm p},n_\mathrm{r}}$,
the received signal in \eqref{eq::cs0} becomes:
\begin{equation}
\begin{aligned}[b]
\mathbf{y}_{\mathrm{u},m, k} &= \mathbf{H}_k^\mathrm{T} \sum_r \alpha_{m, k, n_\mathrm{r}} {\bf s}_{{\rm p},n_\mathrm{r}} + \mathbf{z}_{\mathrm{u},m, k}.
\end{aligned}
\end{equation}
In this way, one may separately estimate the $n_\mathrm{r}$-th SIMO channel $[\mathbf{H}_1^\mathrm{T} {\bf s}_{{\rm p},n_\mathrm{r}},\mathbf{H}_2^\mathrm{T} {\bf s}_{{\rm p},n_\mathrm{r}},\cdots,\mathbf{H}_K^\mathrm{T} {\bf s}_{{\rm p},n_\mathrm{r}}]^\mathrm{T}$ for $n_\mathrm{r}=\{1,2,\cdots,N_\mathrm{r}\}$. Then, $\mathbf{H}_k$ can be reconstructed by removing ${\bf S}_{\rm p}$ from $\mathbf{H}_k^\mathrm{T} {\bf S}_{\rm p}$.
We denote the collection of pilot symbols as $\mathbfcal{Y}_\mathrm{u} = \{\{\mathbf{y}_{\mathrm{u},m, k}\}_{m=1}^{M}\}_{k=1}^{K}$.

\subsection{{Modulation and Precoding at TX}}
Following CSIT estimation, symbol modulation and precoding are performed for downlink transmission. For symbol modulation, we consider a QAM with order $2^B$. The QAM modulator maps $B$ bits to a complex symbol via amplitude modulation of the in-phase and the quadrature phase component, which is expressed as:
\begin{equation}
r_{m, k, \ell}=\mathrm{g}_{\mathrm{q},B}({\bm \beta}_{m,k,\ell}),
\end{equation}
where ${\bm \beta}_{m,k,\ell} = [{\beta}_{m,k,\ell,1}, \cdots,{\beta}_{m,k,\ell,B}]^{\rm T}$, ${\beta}_{m,k,\ell,b} \in \{-1,1\}$ denotes the bit sequence to be transmitted in $k$-th sub-carrier and $\ell$-th spatial stream of the $m$-th OFDM symbol, and $r_{m, k, \ell}$ is the modulated QAM symbol. We denote the collection of all the transmitted bits as $\boldsymbol{\beta} = \{\{\{{\bm \beta}_{m,k,\ell}\}_{m=1}^{M}\}_{k=1}^{K}\}_{\ell=1}^{L}$.
Let $\mathbf{r}_{m,k} = [r_{m,k, 1}, \cdots, r_{m,k, L} ]^{\rm T}\in \mathbb{C}^{L}$, which is further mapped to the transmit antenna via precoding.
The received payload is given by
\begin{equation}
\begin{aligned}[b]\label{eq::sys_ydata}
\mathbf{y}_{\mathrm{d}, m, k} = \mathbf{H}_{k}  \mathbf{V}_k \mathbf{r}_{m,k} + \mathbf{z}_{\mathrm{d}, m, k},
\end{aligned}
\end{equation}
where $\mathbf{V}_k \in \mathbb{C}^{N_\mathrm{t} \times L}$ denotes the precoder at the $k$-th subcarrier, and $\mathbf{z}_{\mathrm{d}, m, k}\sim \mathcal{CN}(0, \sigma_{\rm r}^2 {\bf I}_K)$ is the AWGN. We denote the collection of downlink data observations as $\mathbfcal{Y}_\mathrm{d} = \{\{\mathbf{y}_{\mathrm{d},m, k}\}_{m=1}^{M}\}_{k=1}^{K}$.

\subsection{{Downlink DMRS, Decorrelator, and Demodulation}}\label{sec:DMRS_demod}
To recover $\mathbf{r}_{m,k}$ from $\mathbf{y}_{\mathrm{d}, m, k}$, the primary task is to estimate the CSIR, which is the downlink effective channel given by:
\begin{equation}\label{eq::effch}
\mathbf{H}_{\mathrm{e}, k} = \mathbf{H}_k \mathbf{V}_k,
\end{equation}
for each subcarrier. To achieve this, a straightforward solution is to transmit DMRS over $L$ OFDM symbols, allowing the RX to observe:
\begin{equation}
\begin{aligned}[b]\label{eq::ydmrs}
\mathbf{Y}_{\mathrm{p}, k}
&= \mathbf{H}_{\mathrm{e}, k}{\bf R}_{\rm p} + \mathbf{Z}_{\mathrm{p}, k},
\end{aligned}
\end{equation}
where ${\bf R}_{\rm p}\in \mathbb{C}^{L \times L} $ is an unitary matrix. The channel estimate $\mathbf{H}_{\mathrm{e}, k}$ can then be obtained using a linear minimum mean square error (LMMSE) estimator:
\begin{equation}
\hat{\mathbf{H}}_{\mathrm{e}, k} = \frac{1}{1 + \sigma_{\rm r}^2}\mathbf{Y}_{\mathrm{p}, k} {\bf R}_{\rm p}^{\rm H}.
\end{equation}
Next, the LMMSE decorrelator for the $k$-th subcarrier is derived using $\hat{\mathbf{H}}_{\mathrm{e}, k}$, as follows:
\begin{equation}
\begin{aligned}[b]\label{eq::lmmse_u}
{\bf U}_k = \hat{\mathbf{H}}_{\mathrm{e}, k} (\hat{\mathbf{H}}_{\mathrm{e}, k}^\mathrm{H} \hat{\mathbf{H}}_{\mathrm{e}, k} + \sigma_{\rm r}^2 \mathbf{I}_\mathrm{L})^{-1},
\end{aligned}
\end{equation}
and the received data streams become
\begin{equation}
\begin{aligned}[b]\label{eq::sys_rk}
\hat{\bf r}_{m,k} = {\bf U}_k^\mathrm{H} \mathbf{y}_{\mathrm{d}, m, k}.
\end{aligned}
\end{equation}

Finally, the RX demodulates $\hat{\bf r}_{m,k}=[\hat{r}_{m,k,1},\cdots,\hat{r}_{m,k,L}]^{\rm T}$ to estimate the probability ${\bm \rho}_{m,k,\ell}= [{\rho}_{m,k,\ell,1}, \cdots,{\rho}_{m,k,\ell,B}]^{\rm T}$:
\begin{equation}\label{eq::demodulate}
{\bm \rho}_{m,k,\ell} = \mathrm{g}^{-1}_{\mathrm{q},B}(\hat{r}_{m,k,\ell}),
\end{equation}
where ${ \rho}_{m,k,\ell,b} \in [0,1]$ is the estimated probability for $\p({\beta}_{m,k,\ell,b} = 1)$, and
$\mathrm{g}^{-1}_{\mathrm{q},B}(\cdot)$ is an NN-based QAM de-mapper introduced in \cite{hoydis2022sionna} which can be trained separately. The cross entropy loss is given by:
\begin{equation}\begin{aligned}[b]
\mathcal{L}_\mathrm{ce}&({\beta}_{m,k,\ell,b},\hat{\rho}_{m,k,\ell,b}) =\\& {\beta}_{m,k,\ell,b} \log(\hat{\rho}_{m,k,\ell,b}) + (1-{\beta}_{m,k,\ell,b}) \log(1-\hat{\rho}_{m,k,\ell,b}),
\end{aligned}\end{equation}
During inference, we simply apply hard decision on ${\bm \rho}_{m,k,\ell}$:
\begin{equation}
\hat{\bm \beta}_{m,k,\ell} = \mathbbm{1}(\hat{\bm \rho}_{m,k,\ell} >0).
\end{equation}
The  BER is given by:
\begin{equation}
\psi=\frac{1}{2 MKLB}\sum_{m,k,\ell,b}\left(1-{\beta}_{m,k,\ell,b} \hat{\beta}_{m,k,\ell,b}\right).
\end{equation}

\section{Optimization-Based Approach for Delay-Domain Sparse Precoding to Reduce DMRS Overhead}
As discussed in Section \ref{sec:DMRS_demod}, the conventional DMRS design necessitates $L$ OFDM pilot symbols, which can be expensive when the number of streams (i.e., $L$) is large. In this section, we propose an optimization-based method for designing the precoder to create a sparse effective channel, facilitating the use of the FDM pilot scheme introduced in Section \ref{sec:SRS_signal}.

\subsection{Concerns in Conventional Precoder Design}
\subsubsection{Non-Sparsity of the Effective Channel}
We first explain why the FDM pilot scheme for SRS cannot be directly applied to DMRS. We denote the stacked effective channel as $\mathbf{\overline{H}}_{\rm e} = [\mathbf{H}_{\mathrm{e}, 1}^{\rm T}, \cdots, \mathbf{H}_{\mathrm{e}, K}^{\rm T}]^{\rm T} \in \mathbb{C}^{K N_{\rm r} \times  L}$, and the stacked precoder as $\overline{\bf V}=[{\bf V}_1^{\rm T}, \cdots, {\bf V}_K^{\rm T}]^T \in \mathbb{C}^{K N_{\rm t} \times L}$. Similar to the propagation channel, we define the stacked effective channel in the delay domain as follows:
\begin{equation}\label{eq::X_eff}
\overline{\mathbf{X}}_{\rm e} = (\mathbf{F}^\mathrm{H} \otimes \mathbf{I}_{N_\mathrm{t}}) \mathbf{\overline{H}}_{\rm e},
\end{equation}
and the precoder $\overline{\bf V}$ in delay domain is defined by:
\begin{equation}\label{eq::W_define}
\overline{\mathbf{W}} = (\mathbf{F}^\mathrm{H}\otimes \mathbf{I}_{N_\mathrm{t}}) \mathbf{\overline{V}}.
\end{equation}

To further elaborate on delay-domain sparsity, we define a selection vector $\mathbf{e}_{I}^{(i)} \in \mathbb{R}^{I}$ as follows:
\begin{equation}\label{eq::selection_e}
\mathbf{e}_{I}^{(i)} = [0,\cdots,1,\cdots,0]^\mathrm{T},
\end{equation}
where all elements are zeros except for the $i$-th element, which is set to one. This leads to the antenna selection matrix for the stacked variables, defined as follows:
\begin{equation}\label{eq::selection_E_antenna}
\mathbf{E}_{{\rm A},N}^{(n)} = \mathbf{I}_K \otimes \mathbf{e}_{N}^{(n)}.
\end{equation}
Using \eqref{eq::selection_e} and \eqref{eq::selection_E_antenna}, we can express the elements in the stacked variables related to the $n_\mathrm{r}$-th receive antenna, $n_\mathrm{t}$-th transmit antenna, and the $\ell$-th data stream as follows:
\begin{align}
\mathbf{\overline{h}}_{n_\mathrm{r}, n_\mathrm{t}} &= \mathbf{E}_{{\rm A},{N_\mathrm{r}}}^{({n_\mathrm{r}})} \mathbf{\overline{H}}_{\rm e} \mathbf{e}_{N_\mathrm{t}}^{(n_\mathrm{t})},\\
\mathbf{\overline{h}}_{\mathrm{e}, n_\mathrm{r}, \ell} &= \mathbf{E}_{{\rm A},{N_\mathrm{r}}}^{({n_\mathrm{r}})} \mathbf{\overline{H}}_{\rm e} \mathbf{e}_{L}^{(\ell)}, \\
\overline{\mathbf{v}}_{n_\mathrm{t}, \ell} &= \mathbf{E}_{{\rm A},{N_\mathrm{t}}}^{({n_\mathrm{t}})} \mathbf{\overline{V}} \mathbf{e}_{L}^{(\ell)},\\
\mathbf{\overline{x}}_{n_\mathrm{r}, n_\mathrm{t}} &= \mathbf{E}_{{\rm A},{N_\mathrm{t}}}^{({n_\mathrm{t}})} \mathbf{\overline{X}} \mathbf{e}_{N_\mathrm{t}}^{(n_\mathrm{t})},\\
\mathbf{\overline{x}}_{\mathrm{e}, n_\mathrm{r}, \ell} &= \mathbf{E}_{{\rm A},{N_\mathrm{r}}}^{({n_\mathrm{r}})} \mathbf{\overline{X}}_{\rm e} \mathbf{e}_{L}^{(\ell)},\\
\overline{\mathbf{w}}_{n_\mathrm{t}, \ell} &= \mathbf{E}_{{\rm A},{N_\mathrm{t}}}^{({n_\mathrm{t}})} \mathbf{\overline{W}} \mathbf{e}_{L}^{(\ell)}.
\end{align}

From \eqref{eq::effch}, the relationship among $\mathbf{\overline{h}}_{n_\mathrm{r}, n_\mathrm{t}}$, $\mathbf{\overline{h}}_{\mathrm{e}, n_\mathrm{r}, \ell}$,  and $\overline{\mathbf{v}}_{n_\mathrm{t}, \ell}$ is expressed as:
\begin{equation}\label{eq::h_eff_new}
\mathbf{\overline{h}}_{\mathrm{e},  n_\mathrm{r}, \ell} = \sum_{n_\mathrm{t}=1}^{N_\mathrm{t}} \mathbf{\overline{h}}_{n_\mathrm{r}, n_\mathrm{t}}  \odot \overline{\mathbf{v}}_{n_\mathrm{t}, \ell} ,
\end{equation}
where $\odot$ indicates the Hadamard product. The product relation translates to convolution in the delay domain, expressed as:
\begin{equation}
\begin{aligned}[b]\label{eq::x_eff_nr_nt}
\mathbf{\overline{x}}_{\mathrm{e}, n_\mathrm{r}, \ell}
&= \sum_{n_\mathrm{t}=1}^{N_\mathrm{t}}  \mathbf{\overline{x}}_{n_\mathrm{r}, n_\mathrm{t}}  \ast \overline{\mathbf{w}}_{n_\mathrm{t}, \ell},
\end{aligned}
\end{equation}
where $\ast$ indicates the convolution operation. Therefore, the sparsity of $\mathbf{\overline{x}}_{\mathrm{e}, n_\mathrm{r}, \ell}$ depends on both $\mathbf{\overline{x}}_{n_\mathrm{r}, n_\mathrm{t}}$ and $\overline{\mathbf{w}}_{n_\mathrm{t}, \ell}$ for all $n_\mathrm{t}$. We have illustrated the sparsity of $\mathbf{\overline{x}}_{n_\mathrm{r}, n_\mathrm{t}}$ in \Figref{fig::sparsity}{}. However, the delay-domain precoder $\overline{\mathbf{w}}_{n_\mathrm{t}, \ell}$ is typically not sparse.


\begin{figure}[!t]
    \centering
        \subfigure[Delay domain original channel $\mathbf{\overline{x}}_{1, n_\mathrm{t}}, n_\mathrm{t}=1$]    {
	    	\includegraphics[trim={0 0 0 0},clip,width=0.7\linewidth]{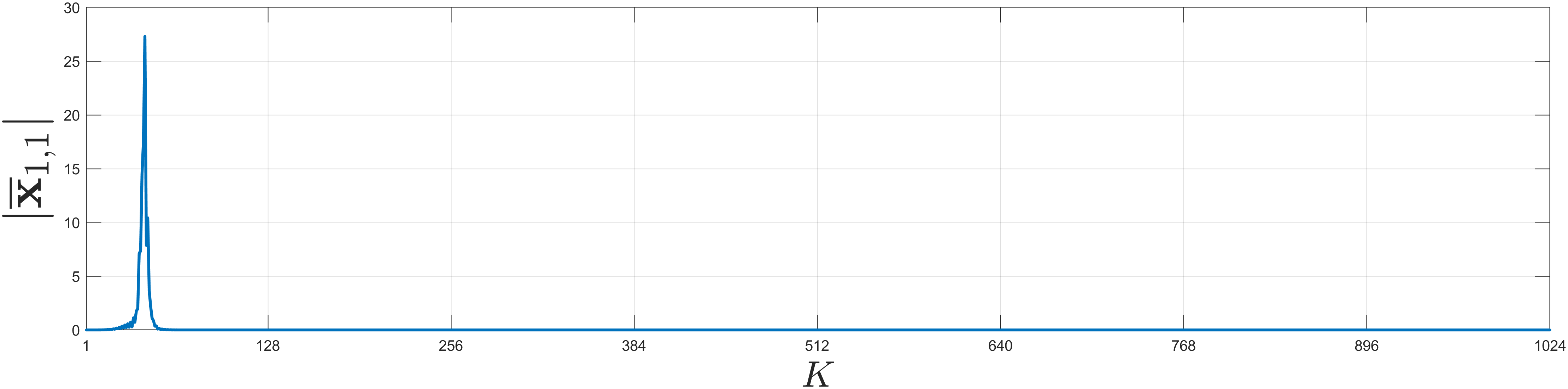}
	    	}
        \subfigure[Delay domain original channel $\mathbf{\overline{x}}_{1, n_\mathrm{t}}, n_\mathrm{t}=2$]    {
	    	\includegraphics[trim={0 0 0 0},clip,width=0.7\linewidth]{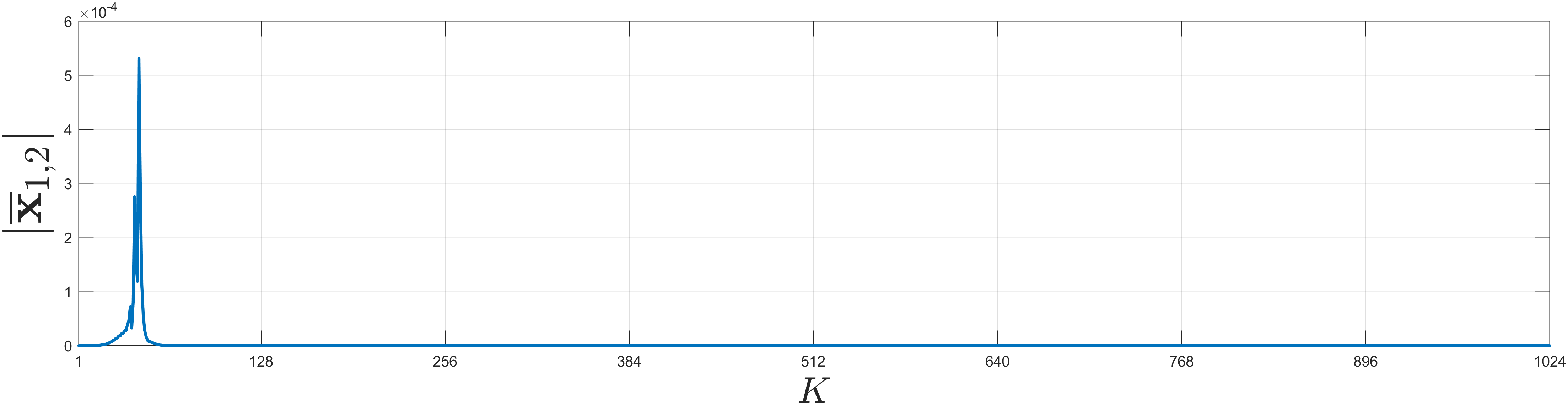}
	    	}
    \subfigure[Delay domain precoder, $\overline{\mathbf{w}}_{n_\mathrm{t}, 1}, n_\mathrm{t}=1$]    {
	    	\includegraphics[trim={0 0 0 0},clip,width=0.7\linewidth]{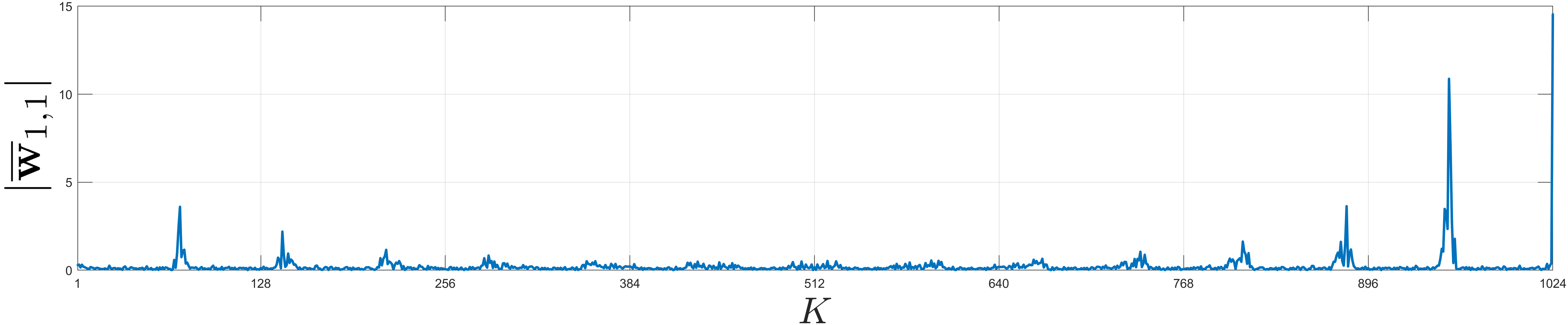}
	    	}
	    \subfigure[Delay domain precoder, $\overline{\mathbf{w}}_{n_\mathrm{t}, 1}, n_\mathrm{t}=2$]    {
	    	\includegraphics[trim={0 0 0 0},clip,width=0.7\linewidth]{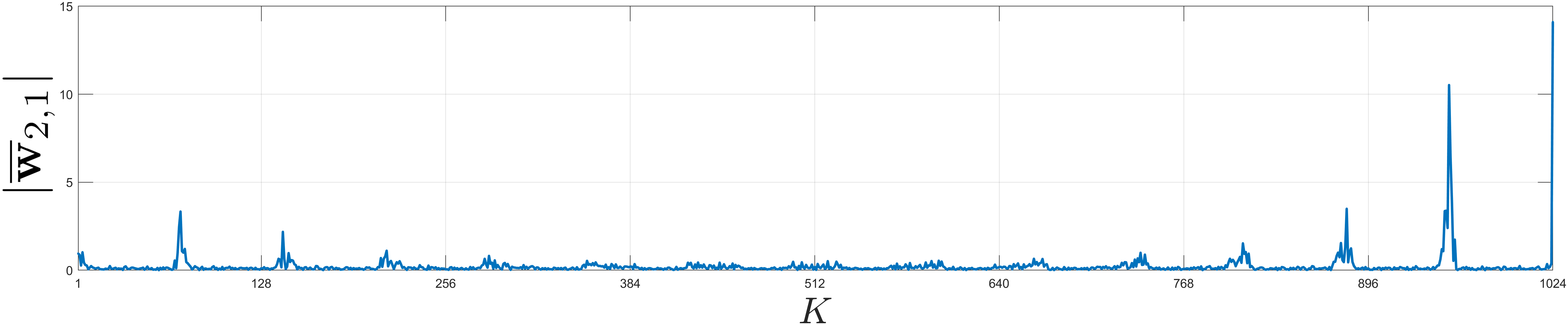}
		}
    \subfigure[Delay domain effective channel $\mathbf{\overline{x}}_{\mathrm{e},  1, 1}$]    {
	    	\includegraphics[trim={0 0 0 0},clip,width=0.7\linewidth]{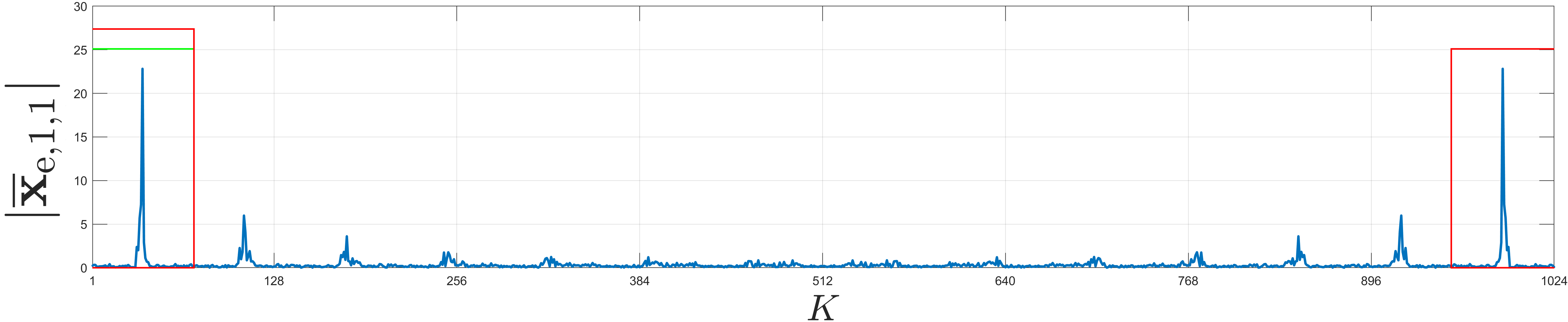}
	    	}
	\caption{Illustration of the delay domain channel, precoder and the effective channel with $N_\mathrm{r}=1$, $N_\mathrm{t}=2$, $L=1$, and $K=1024$ for the 802.11ax MIMO channel with the delay profile configured to model B. }\label{fig::sparsity-eff}
\end{figure}

We elaborate on this issue with a toy example, as illustrated in \Figref{fig::sparsity-eff}{}. Specifically, we consider a MISO-OFDM transmission with  $N_\mathrm{r}=1$, $N_\mathrm{t}=2$, $L=1$, and $K=1024$, where the precoder ${\bf V}_k$ is designed using an SVD-based scheme \cite{perahia2013next, ali2017beamforming, tiwari2023advancing} for each subcarrier separately.  {\Figref{fig::sparsity-eff}(a)} to {\Figref{fig::sparsity-eff}(d)} illustrate the sparsity of  $\mathbf{\overline{x}}_{1, 1}$, $\mathbf{\overline{x}}_{1, 2}$, $\overline{\mathbf{w}}_{1, 1}$ and $\overline{\mathbf{w}}_{2, 1}$ , respectively, while {\Figref{fig::sparsity-eff}(e)} illustrates the sparsity of $\mathbf{\overline{x}}_{\mathrm{e}, 1, 1}$. It is evident that $\overline{\mathbf{w}}_{1, 1}$ and $\overline{\mathbf{w}}_{2, 1}$ contain numerous peaks in the delay domain, resulting in $\mathbf{\overline{x}}_{\mathrm{e}, 1, 1}$ not being sparse. In particular, applying an anti-aliasing filter to select the first 72 delay taps of $\mathbf{\overline{x}}_{\mathrm{e},  1, 1}$ yields an NMSE of 0.593, which is infeasible for effective decorrelation and demodulation.  Even when we increase the filter window to select channel responses in both the first and last 72 delay taps, the NMSE remains at 0.186, which is still unacceptable for good performance.

\subsubsection{Performance Metric on Achievable Rate is not Efficient}
\begin{figure}
    \centering
	    	\includegraphics[trim={0 0 0 0},clip,width=0.7\linewidth]{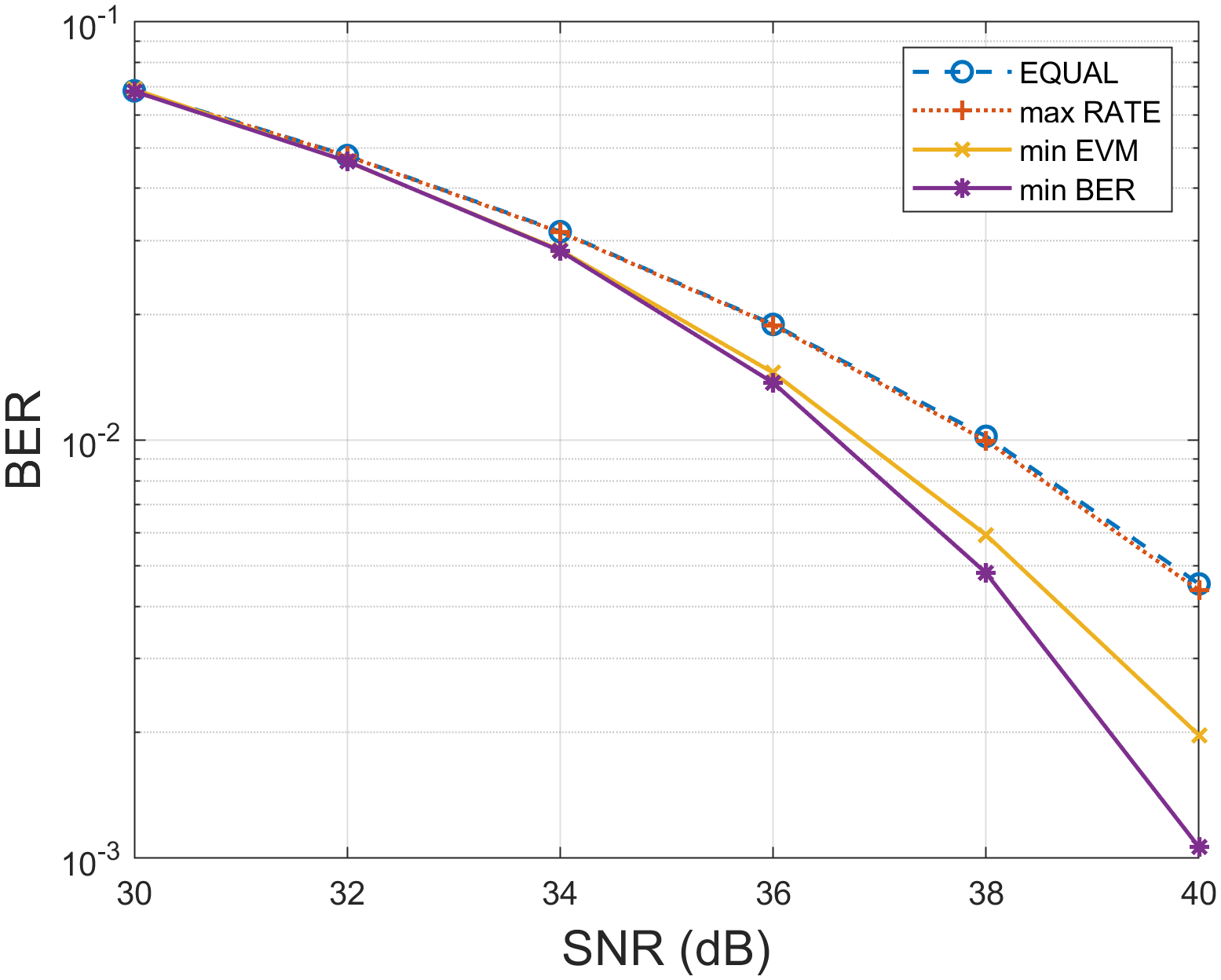}
	\caption{Illustration of BER vs SNR with $N_\mathrm{t}=2$, $N_\mathrm{r}=1$, $L=1$ and $K=1024$ for 4096 QAM modulation over the 802.11ax  MIMO channel, with the delay profile configured to model B.}\label{fig::ber-vs-snr-init}
\end{figure}

Most existing works optimize the precoder primarily to maximize the achievable rate. However, these designs often result in suboptimal performance in practice, as all subcarriers assigned to the same receiver typically utilize the same MCS. To illustrate this issue, we consider the channel configuration shown in \Figref{fig::sparsity-eff}{}. We optimize power allocation across subcarriers, focusing on performance metrics such as rate, EVM, and BER, as depicted in \Figref{fig::ber-vs-snr-init}. The rate metric shows only slight improvement compared to the equal-power allocation baseline. While the EVM metric performs better, it remains suboptimal at high SNR. The BER is the most effective metric as it reflects end-to-end performance. However, designing algorithms for this metric becomes challenging when additional constraints, such as the delay-domain sparsity constraint introduced in the next subsection, are included.

%
%


\subsection{Proposed Delay-Domain Sparse Precoder for EVM Minimization}\label{sec::opt_sparse_precoder}
\subsubsection{Common Precoder for All Subcarriers}\label{sec::comm_sparse_precoder}
A straightforward method to ensure delay domain sparsity of the effective channel is to utilize a common precoder across all sub-carriers given by the first $L$ right singular vectors of the sample covariance $\mathbf{C}_\mathrm{h}$, which is given by
\begin{equation}
\mathbf{C}_\mathrm{h} = \sum_{k=1}^{K} \mathbf{H}_k^\mathrm{H} \mathbf{H}_k.
\end{equation}
This results in a precoder with a single tap in the delay domain and hence, the sparsity feature of the effective channel is same as that of the propagation channel.  However, this precoder design suffers performance degradation since the OFDM channels are frequency selective.

\subsubsection{Sparse Delay Domain Precoder}
A more effective way to maintain a sparse delay-domain precoder ($\overline{\bf W}$) while maintaining good performance is to apply a low-pass filter in the delay domain, analogous to anti-aliasing filter, resulting in:
\begin{equation}\label{eq::delay_W}
\overline{\mathbf{V}} = (\widetilde{\mathbf{F}}_\mathrm{v} \otimes \mathbf{I}_{N_\mathrm{t}}) \mathbf{\widetilde{W}},
\end{equation}
where $\widetilde{\mathbf{F}}_\mathrm{v} \in \mathbb{C}^{K \times D_{\rm v}}$ is the partial DFT matrix by truncating $\mathbf F$ to retain only the first $D_{\rm v}$ columns, and $\mathbf{\widetilde{W}} \in \mathbb{C}^{D_\mathrm{v}  N_\mathrm{t} \times L}$ is the sparse delay domain precoder obtained by retaining the first $D_\mathrm{v}  N_\mathrm{t}$ rows. Similar to \eqref{eq::selection_E_antenna}, we define the subcarrier selection matrix for the stacked variables by $\mathbf{E}_{{\rm S},K}^{(k)} = \mathbf{e}_{K}^{(k)} \otimes \mathbf{I}_{N_\mathrm{t}}$. Then, letting $\mathbf{A}_k = \mathbf{E}_{{\rm S},K}^{(k)} (\widetilde{\mathbf{F}}_\mathrm{v} \otimes \mathbf{I}_{N_\mathrm{t}})$, the precoder at the $k$-th subcarrier is represented by a linear function of $\mathbf{\widetilde{W}}$:
\begin{equation}\label{eq::selection_E_subcarrer}
\mathbf{V}_k = \mathbf{A}_k \mathbf{\widetilde{W}}.
\end{equation}

In contrast to \cite{precodedelay2019WCL,zhang2023cross}, we optimize $\mathbf{\widetilde{W}}$ to minimize the EVM, which is give by
\begin{subequations}
\begin{align}
\nonumber
{\mathcal{P}}{(\text{A})}:\,\, & \underset{\mathbf{\widetilde{W}}, \{\mathbf{U}_k\}_{k=1}^{K} }{\mathrm{min}}  \;  \;  \sum_{k=1}^K  \|\mathbf{I}_L - \mathbf{U}_k^\mathrm{H}\mathbf{H}_k\mathbf{V}_k\|_\mathrm{F}^2+\sigma_\mathrm{r}^2\|\mathbf{U}_k\|_\mathrm{F}^2 \\
\label{eq::evm_cons1}
& \;\;\;\;\;\;\quad {\mathrm{s.t.}} \quad \|\mathbf{\widetilde{W}} \|^2_\mathrm{F} \leq K P_{\rm T},
\end{align}
\end{subequations}
where $P_{\rm T}$ is the transmit power budget.
${\mathcal{P}}{(\text{A})}$ is a bi-convex problem which can be solved by iterating:
\begin{itemize}
\item {\bf Updating $\{\mathbf{U}_k\}_{k=1}^{K}$}:
Given precoder $\overline{\mathbf{V}}$, the objective with respect to the decorrelator ${\bf U}_k$ is decoupled, and optimal solution can be derived with closed form:
\begin{equation}\label{eq::uk}
\mathbf{U}_k = \mathbf{H}_k\mathbf{V}_k ((\mathbf{H}_k\mathbf{V}_k)^\mathrm{H} \mathbf{H}_k\mathbf{V}_k + \sigma_\mathrm{r}^2 \mathbf{I}_{L})^{-1}.
\end{equation}

\item {\bf Updating $\mathbf{\widetilde{W}}$}:
Given decorrelator, optimal solution for $\mathbf{\widetilde{W}}$ can be derived as follows:
\begin{equation}\begin{aligned}[b]\label{eq::update_w}
\mathbf{\widetilde{W}} &= \left(\sum_k \mathbf{A}_k^\mathrm{H}\mathbf{H}_\mathrm{k}^\mathrm{H} \mathbf{U}_k \mathbf{U}_k^\mathrm{H} \mathbf{H}_\mathrm{k} \mathbf{A}_k + \upsilon \mathbf{I}_{D_\mathrm{v} N_\mathrm{t}}\right)^{-1}\\&
\quad\quad\quad\quad\quad\quad\quad\quad\quad\quad\quad \left( \sum_k \mathbf{A}_k^\mathrm{H}\mathbf{H}_\mathrm{k}^\mathrm{H} \mathbf{U}_k\right),
\end{aligned}\end{equation}
where $\upsilon$ is the Lagrange multiplier corresponding to the power constraint which can be obtained using bisection search.
\end{itemize}

\begin{figure}[!t]
    \centering
	    	\includegraphics[trim={0 0 0 0},clip,width=0.9\linewidth]{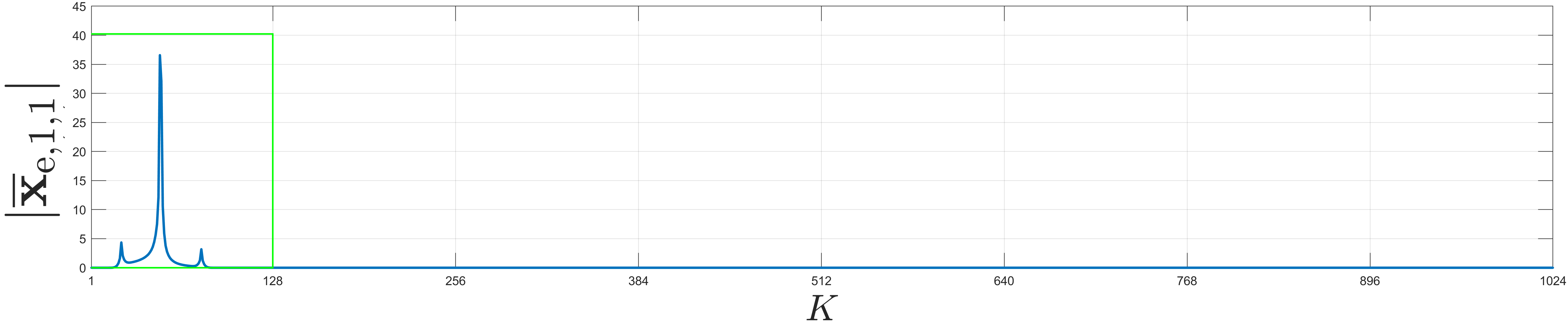}
	\caption{Illustration on the sparsity of the effective channel for the solution of ${\mathcal{P}}{(\text{A})}$.}\label{fig::sparsity-eff-pre}
\end{figure}

\subsection{Reducing DMRS Overhead by Exploiting Delay-Domain Sparsity}
The proposed solution above produces a delay-domain sparse channel with an effective delay spread of:
\begin{equation}
D_\mathrm{eff} = D + D_\mathrm{v} - 1
\end{equation}
We illustrate the effective channel in \Figref{fig::sparsity-eff-pre}{} with $D_\mathrm{v}=56$, using the same channel configuration as in \Figref{fig::sparsity-eff}{}. The effective channel can be perfectly reconstructed when the window size of the anti-aliasing filter exceeds 128. Therefore, the FDM pilot scheme becomes applicable for DMRS and costs $\overline{M} = \lceil \frac{L D_\mathrm{eff}}{K}\rceil$ OFDM symbols. Analogous to uplink pilot allocation, we can allocate upto $A_\mathrm{r} = \lfloor K/D_\mathrm{eff} \rfloor$ simultaneous pilots over one OFDM symbol. 
Similar to \ref{sec::mimo_fdm}, we consider a unitary matrix ${\bf \overline{S}}_{\rm p}=[{\bf \overline{s}}_{{\rm p},1},{\bf \overline{s}}_{{\rm p},2},\cdots,{\bf \overline{s}}_{{\rm p},L}]^\mathrm{T} \in \mathbb{C}^{L \times L}$. We can map ${\bf \overline{s}}_{{\rm p},\ell}$ to the  $L$ orthogonal pilot sequences using the following indicator:
\begin{equation}
\overline{\alpha}_{\overline{m}, k,  \ell} = \begin{cases}
				1,& {\rm{if}}\; \lceil \frac{\ell}{A_\mathrm{r}}\rceil=\overline{m} \; {\rm{and}} \,\,\, k \in \Omega_{\ell\%A_\mathrm{r}},\\
				0,&  \mathrm{otherwise}.
			\end{cases}
\end{equation}
Then, letting $\mathbf{\overline{s}}_{\mathrm{u}, \overline{m}, k }=\sum_\ell \overline{\alpha}_{k, \overline{m}, \ell} {\bf \overline{s}}_{{\rm p},\ell}$,
the received signal in \eqref{eq::ydmrs} becomes:
\begin{equation}\label{eq::sparse_dmrs}
\begin{aligned}[b]
\mathbf{y}_{\mathrm{p},m, k} &= \mathbf{H}_{\mathrm{e},k}^\mathrm{T} \sum_r \overline{\alpha}_{\overline{m}, k,  \ell} {\bf \overline{s}}_{{\rm p},\ell} + \mathbf{z}_{\mathrm{p},\overline{m}, k}.
\end{aligned}
\end{equation}
The collection of downlink pilot observations is denoted as $\mathbfcal{Y}_\mathrm{p} = \{\{\mathbf{y}_{\mathrm{p},\overline{m}, k}\}_{\overline{m}=1}^{\overline{M}}\}_{k=1}^{K}$.




\section{Datapath Design of the Proposed End-to-end DNN for MIMO-OFDM System}\label{sec:detapath}

This section illustrates the end-to-end DNN data path, which is divided into three key modules: the propagation channel estimation module, the precoder design module, and the effective channel estimation module. Together, these modules enable the end-to-end optimization of the framework.

\begin{figure*}[!t]
    \centering
	    	\includegraphics[trim={0 0 0 0},clip,width=0.9\linewidth]{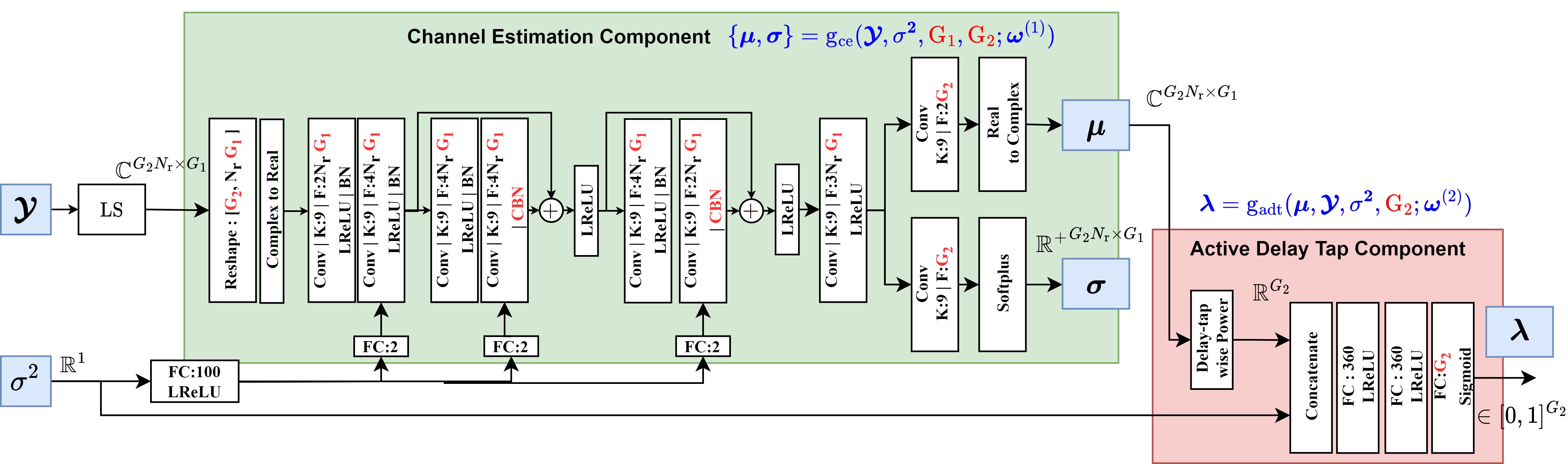}
	\caption{Architecture for channel estimation and active delay taps estimation in the propagation channel estimation module and the effective channel estimation module. The architecture is parameterized by weights $\{{\bm \omega}^{(1)}, {\bm \omega}^{(2)}\}$ providing the mean, variance and a probability that a given delay tap is active which aids the Bayesian training. The architecture contains configurable parameters ${G}_1$ and ${G}_2$. 
The $\mathrm{FC}:\mathrm{\overline{K}}$  represents the fully connected layers with $\overline{K}$ output nodes, $\mathrm{Conv}|\mathrm{K}:\overline{K}|\mathrm{F}:\overline{F}$ indicates the 1D convolution layer with $\overline{F}$ filters of kernel size $\overline{K}$. CBN stands for conditional batch normalization \cite{guo2023robust}, which is used to adapt to varying noise variances.
}\label{fig::module:prop:key}
\end{figure*}

\begin{figure*}[!t]
    \centering
	    	\includegraphics[trim={0 0 0 0},clip,width=0.8\linewidth]{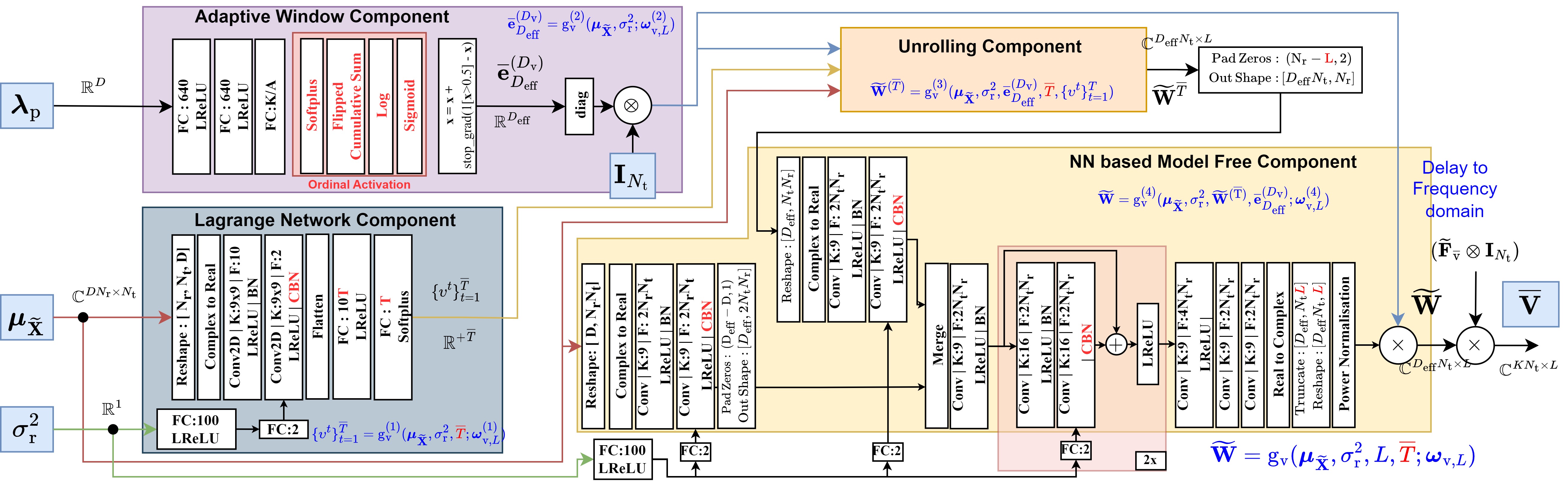}
	\caption{Architecture of the Precoding Design Module implementing the algorithmic unrolling with NN based initialization and Lagrange network. Here $\mathrm{Conv}|\mathrm{K}:\overline{K}_1\times \overline{K}_2|\mathrm{F}:\overline{F})$ is the 2D convolution layer with $\overline{F}$ filters of kernel size $\overline{K}_1 \times \overline{K}_2$. A novel adaptable widow component adjusts the delay spread of the final precoder output. The Unrolling component iteratively updates the decorrelator and the delay domain precoder using \eqref{eq::uk} and \eqref{eq::update_w}.}\label{fig::module:prec:key}
\end{figure*}

\begin{figure}[!t]
    \centering
	    	\includegraphics[trim={0 0 0 0},clip,width=0.8\linewidth]{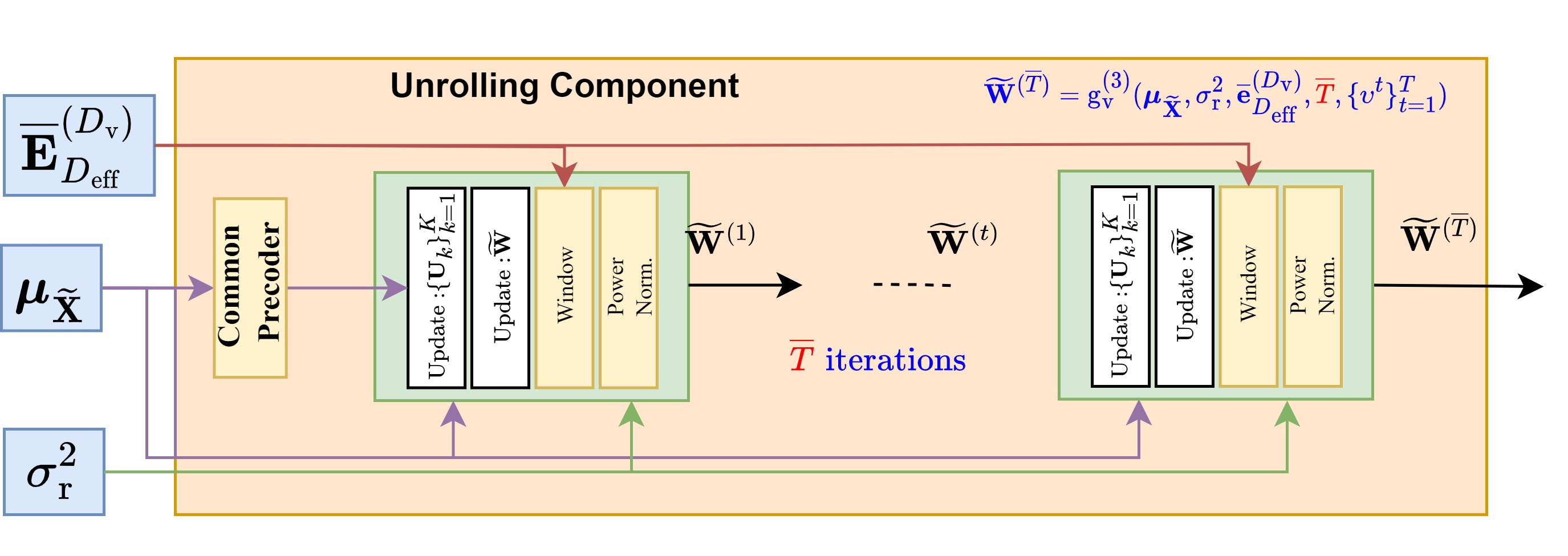}
	\caption{Architecture of the Unrolling component initializer for the precoder design using \eqref{eq::uk} and \eqref{eq::update_w} along with explicit window and power normalization.}\label{fig::module:prec:unroll}
\end{figure}

\subsection{NN-based Channel and Active Delay Tap Estimation}
We first introduce the NN based channel estimation (NN-CE) and active delay tap estimation (NN-ADT) component illustrated in \Figref{fig::module:prop:key}{}, represented by the following functions:
\begin{align}
\label{eq::nnce_1}
\{\boldsymbol{\mu}, \boldsymbol{\sigma} \} &= \mathrm{g}_\mathrm{ce}(\mathbfcal{Y}, \mathbf{\sigma}^2, {G}_1, {G}_2; {\bm \omega}^{(1)}),\\
\label{eq::nnce_2}
\boldsymbol{\lambda} &= \mathrm{g}_\mathrm{adt}(\boldsymbol{\mu}, \mathbfcal{Y}, \mathbf{\sigma}^2, {G}_2; {\bm \omega}^{(2)}),
\end{align}
where $\{\boldsymbol{\mu}, \boldsymbol{\sigma} \} \in \{\mathbb{C}^{G_2 N_\mathrm{r} \times G_1}, {\mathbb{R}^+}^{G _2N_\mathrm{r} \times G_1} \}$ are the channel mean and element-wise variances, $\boldsymbol{\lambda}\in [0,1]^{G_2}$ are the active delay tap estimates, ${G}_1$ and  ${G}_2$ are the architecture parameters,
and ${\bm \omega}^{(1)}$ and ${\bm \omega}^{(2)}$ are the weights of the channel estimation and active delay component.

These two components are designed by leveraging the characteristics of delay-domain sparsity, which will be utilized by both the propagation channel estimation module and the effective channel estimation module:
\subsubsection{Propagation Channel Estimation Module}
This module is deployed at the TX and takes as input the SRS $\mathbfcal{Y}_\mathrm{u}$ and the uplink noise variance $\sigma_\mathrm{u}^2$. It provides estimates of the mean, variance, and the active delay taps of the CSIT.
\begin{itemize}
\item{\textbf{Propagation Channel Estimation Component}:} This component is defined by substituting ${G}_1= N_\mathrm{t}$ and  ${G}_2=D$ into the NN-CE function in \eqref{eq::nnce_1}:
\begin{equation}\begin{aligned}[b]
\label{eq::datapath_prop_mu_sigma}
\{\boldsymbol{\mu}_{\mathbf{\widetilde{X}}}, \boldsymbol{\sigma}_{\mathbf{\widetilde{X}}} \}
&= \mathrm{g}^{(1)}_\mathrm{srs}(\mathbfcal{Y}_\mathrm{u}, \sigma_\mathrm{u}^2; {\bm \omega}_\mathrm{h}^{(1)})\\
&= \mathrm{g}_\mathrm{ce}(\mathbfcal{Y}_\mathrm{u}, \sigma_\mathrm{u}^2,  N_\mathrm{t}, D; {\bm \omega}_\mathrm{h}^{(1)}).
\end{aligned}\end{equation}

\item{\textbf{Propagation Active Delay Component}:} Substituting ${G}_2=D$ into the NN-ADT function in \eqref{eq::nnce_2}, we have
\begin{equation}\begin{aligned}[b]\label{eq::datapath_prop_lam}
\boldsymbol{\lambda}_\mathrm{h}
&= \mathrm{g}^{(2)}_\mathrm{srs}(\boldsymbol{\mu}_{\mathbf{\widetilde{X}}}, \mathbfcal{Y}_\mathrm{u}, \sigma_{\rm u}^2; {\bm \omega}_\mathrm{h}^{(2)})\\
&= \mathrm{g}_\mathrm{adt}(\boldsymbol{\mu}_{\mathbf{\widetilde{X}}}, \mathbfcal{Y}_\mathrm{u}, \sigma_{\rm u}^2, D; {\bm \omega}_\mathrm{h}^{(2)}).
\end{aligned}\end{equation}
\end{itemize}

We summarize the collection of weights in the propagation channel estimation module as ${\bm \omega}_\mathrm{h} = \{{\bm \omega}_\mathrm{h}^{(1)},{\bm \omega}_\mathrm{h}^{(2)}\}$.

\subsubsection{Effective Channel Estimation Module}
This module is deployed at the RX and takes as input the DMRS $\mathbfcal{Y}_\mathrm{p}$ and the noise variance $\sigma_{0}^2$. It estimates the key variable related to the mean, variance, and the active delay taps of the CSIR.
\begin{itemize}
\item{\textbf{Effective Channel Estimation Component}:} The estimate to the effective channel is obtained as:
\begin{equation}\begin{aligned}[b]
\label{eq::datapath_eff_mu_sigma}
\{\boldsymbol{\mu}_{\mathbf{\widetilde{X}}_\mathrm{e}}, \boldsymbol{\sigma}_{\mathbf{\widetilde{X}}_\mathrm{e}} \}
&= \mathrm{g}^{(1)}_\mathrm{eff}(\mathbfcal{Y}_\mathrm{p}, \sigma_{\rm r}^2; {\bm \omega}_\mathrm{e}^{(1)})\\
&= \mathrm{g}_\mathrm{ce}(\mathbfcal{Y}_\mathrm{p}, \sigma_{\rm r}^2,  L, D_\mathrm{eff}; {\bm \omega}_\mathrm{e}^{(1)}),
\end{aligned}\end{equation}

%

\item{\textbf{Effective Active Delay Component}:} 
Similarly, the active delay tap estimation, $\boldsymbol{\lambda}_\mathrm{e} \in [0,1]^{D_\mathrm{eff}}$, is given as:
\begin{equation}\begin{aligned}[b]\label{eq::datapath_eff_lam}
\boldsymbol{\lambda}_\mathrm{e}
&= \mathrm{g}^{(2)}_\mathrm{eff}(\boldsymbol{\mu}_{\mathbf{\widetilde{X}}_\mathrm{e}}, \mathbfcal{Y}_\mathrm{p}, \sigma_{\rm r}^2; {\bm \omega}_\mathrm{e}^{(2)})\\
&= \mathrm{g}_\mathrm{adt}(\boldsymbol{\mu}_{\mathbf{\widetilde{X}}_\mathrm{e}}, \mathbfcal{Y}_\mathrm{p}, \sigma_{\rm r}^2, D_\mathrm{eff}; {\bm \omega}_\mathrm{e}^{(2)}).
\end{aligned}\end{equation}
\end{itemize}

Collectively, the weights are denoted as ${\bm \omega}_\mathrm{e} = \{{\bm \omega}_\mathrm{e}^{(1)},{\bm \omega}_\mathrm{e}^{(2)}\}$.

\subsection{Precoder design module}
The architecture of the precoder design module is designed using algorithmic unrolling of ${\mathcal{P}}{(\text{A})}$ in Section \ref{sec::opt_sparse_precoder}. As shown in \Figref{fig::module:prec:key}{}, it takes $\boldsymbol{\mu}_{\mathbf{\widetilde{X}}}$ and the noise level $\sigma_\mathrm{r}^2$ as inputs and provides precoder $\mathbf{\widetilde{W}} \in \mathbb{C}^{{D}_\mathrm{v}  N_\mathrm{t} \times L}$:
\begin{equation}\label{eq::precoder_final}
\mathbf{\widetilde{W}} = \mathrm{g}_\mathrm{v}(\boldsymbol{\mu}_{\mathbf{\widetilde{X}}}, \sigma_{\rm r}^2, L, \overline{T}; {\bm \omega}_{\mathrm{v},L}),
\end{equation}
where ${D}_\mathrm{v}$ is the DNN-based precoder delay tap length, ${\bm \omega}_{\mathrm{v},L}$ are the trainable weights for a given number of spatial stream $L$, and $\overline{T}$ denotes the number of unrolled iterations. The function accepts varying values of $L$ without any change in the architecture, enabling architecture re-use for different values of $L$.
The precoder design module specifically consists of 4 components:

\begin{itemize}
\item{\textbf{Lagrange Network Component}:}
Unrolling ${\mathcal{P}}{(\text{A})}$ requires iterative update of $\{\mathbf{U}_k\}_{k=1}^{K} $ and $\mathbf{\widetilde{W}}^{(t)}$ according to \eqref{eq::uk} and \eqref{eq::update_w} for the $t$-th iteration. While both the components have closed form updates, the latter requires obtaining a suitable Lagrangian $\upsilon$ which is time consuming during inference. Inspired by the Lagrange network introduced in \cite{shi2023robust}, where the Lagrangian value is learned through a NN, we propose a Lagrange network component that provides different Lagrange multipliers for all unrolled iterations as follows:
\begin{equation}
\{\upsilon^{t}\}_{t=1}^{\overline{T}} = \mathrm{g}_\mathrm{v}^{(1)}(\boldsymbol{\mu}_{\mathbf{\widetilde{X}}}, \sigma_{\rm r}^2, \overline{T}; {\bm \omega}_{\mathrm{v},L}^{(1)}).
\end{equation}

\item{\textbf{Adaptive Window Component}:}
We achieve a dynamic $1 \leq {D}_\mathrm{v} \leq D_\mathrm{eff}$ for the delay-domain precoder by learning a selection vector:
\begin{equation}\label{eq::prec:supp_e}
\mathbf{\overline{e}}_{D_\mathrm{eff}}^{({D}_{\mathrm{v}})} =
\mathrm{g}_\mathrm{v}^{(2)}(\boldsymbol{\mu}_{\mathbf{\widetilde{X}}}, \sigma_{\rm r}^2; {\bm \omega}_{\mathrm{v},L}^{(2)}),
\end{equation}
$\mathbf{\overline{e}}_{D_\mathrm{eff}}^{(i)} = [1,\cdots,1,\cdots,0]^\mathrm{T},$ is a selection vector of length $D_\mathrm{eff}$ where all elements upto the $i$-th element are 1 while the remaining are set to 0, and ${D}_\mathrm{v}=\|\mathbf{\overline{e}}_{D_\mathrm{eff}}^{({D}_{\mathrm{v}})}\|_0$.
Such structure is ensured by an ordinal activation function as shown in \Figref{fig::module:prec:key}{} cascaded to a hard threshold layer given as follows:
\begin{equation}
f_{\rm{thr}}({\bf x}) = {\bf x} + \mathrm{tf.stop}\_\mathrm{gradient}(\mathbbm{1}({\bf x}>0.5) - {\bf x}).
\end{equation}
Next, we define the selection matrix $\overline{\mathbf{E}}_{{D_\mathrm{eff}}}^{({D}_{\mathrm{v}})}  = \mathrm{diag}(\mathbf{\overline{e}}_{D_\mathrm{eff}}^{({D}_{\mathrm{v}})})\otimes  \mathbf{I}_{N_\mathrm{t}}$, which may constrain the elements in $\mathbf{\widetilde{W}}$ that lie outside the delay-domain window of length ${D}_\mathrm{v}$ to be zero:
\begin{equation}\label{eq::prec:inf:final}
\mathbf{\widetilde{W}}^{(t)}  = \overline{\mathbf{E}}_{{D_\mathrm{eff}}}^{({D}_{\mathrm{v}})}  \mathbf{\widetilde{W}}.
\end{equation}

\item{\textbf{Unrolling Component}:}
As shown in \Figref{fig::module:prec:unroll}{}, the unrolling component takes the initial value and the trainable Lagrange multipliers and unrolls ${\mathcal{P}}{(\text{A})}$ by iteratively updating $\{\mathbf{U}_k\}_{k=1}^{K}$ and $\mathbf{\widetilde{W}}^{(t)}$ according to \eqref{eq::uk} and \eqref{eq::update_w} for ${\overline{T}}$ unrolling iterations:
\begin{equation}
\mathbf{\widetilde{W}}^{(\overline{T})} = \mathrm{g}_\mathrm{v}^{(3)}(\boldsymbol{\mu}_{\mathbf{\widetilde{X}}}, \sigma_{\rm r}^2, \mathbf{\overline{e}}_{D_\mathrm{eff}}^{({D}_{\mathrm{v}})}, {\overline{T}}, \{\upsilon^{t}\}_{t=1}^{T}).
\end{equation}

\item{\textbf{NN-based Model-Free Component}:}
Finally, the unrolling output is forwarded to an NN-based model-free component for additional performance enhancement:
\begin{equation}
\mathbf{\widetilde{W}} = \mathrm{g}_\mathrm{v}^{(4)}(\boldsymbol{\mu}_{\mathbf{\widetilde{X}}}, \sigma_{\rm r}^2, \mathbf{\widetilde{W}}^{(\overline{T})},\mathbf{\overline{e}}_{D_\mathrm{eff}}^{({D}_{\mathrm{v}})}; {\bm \omega}_{\mathrm{v},L}^{(4)}),
\end{equation}
where ${\bm \omega}_\mathrm{v}^{(4)}$ are the weights of the network.
\end{itemize}

Collectively, we denote ${\bm \omega}_{\mathrm{v},L} = \{{\bm \omega}_{\mathrm{v},L}^{(1)}, {\bm \omega}_{\mathrm{v},L}^{(2)}, {\bm \omega}_{\mathrm{v},L}^{(4)}\}$ as the collection of weights in the precoder design module.


\section{Variational Bayesian Training Formulation}
In this section, we introduce the VBI-based training formulation. We first design the generative and inference models  for all key variables in the data path, which are functions of the DNN weights as defined in Section \ref{sec:detapath}. Next, we derive the training loss function by minimizing the KL divergence between the generative and inference models.


\subsection{The Generative Model}\label{sec::gen}

The generative model consists of the likelihood distribution for the observations and the prior distribution that incorporates domain knowledge.
\begin{equation}\label{eq::gen}
\begin{aligned}[b]
&\p_{\rm gen} \bigl(\mathbf{t}_\mathrm{h}, \mathbf{\widetilde{X}},\mathbf{\overline{V}}, \mathbfcal{Y}_\mathrm{u}, \mathbf{t}_\mathrm{e}, \mathbf{\widetilde{X}}_\mathrm{e}, \mathbfcal{Y}_\mathrm{p}, \boldsymbol{\beta}| \mathbfcal{Y}_{\mathrm{d}};{\bm \omega}_\mathrm{pir} \bigr) =
\\
& \p_\mathrm{lik}(\mathbfcal{Y}_\mathrm{u}, \mathbfcal{Y}_\mathrm{p}, \boldsymbol{\beta}| \mathbf{\widetilde{X}},   \mathbf{\widetilde{X}}_\mathrm{e}, \mathbfcal{Y}_{\mathrm{d}}) \p_\mathrm{pri}\bigl(\mathbf{t}_\mathrm{h}, \mathbf{\widetilde{X}},\mathbf{\overline{V}},  \mathbf{t}_\mathrm{e}, \mathbf{\widetilde{X}}_\mathrm{e};{\bm \omega}_\mathrm{pri} \bigr).
\end{aligned}
\end{equation}

\begin{figure}[!t]
    \centering
	    \subfigure[Generative Model]    {
	    	\includegraphics[trim={0 0 0 0},clip,width=0.8\linewidth]{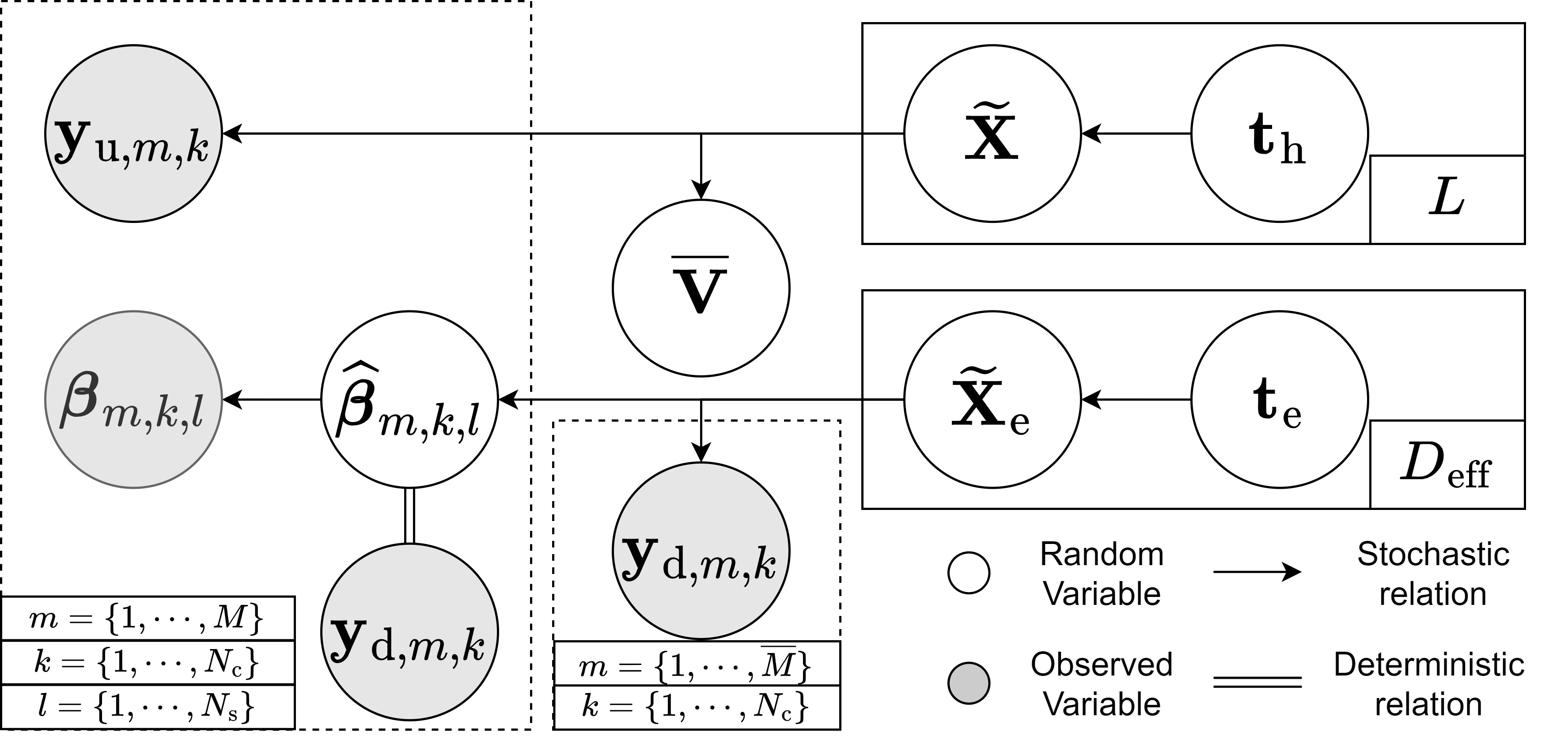}
	    	}
	    \subfigure[Inference Model]    {
	    	\includegraphics[trim={0 0 0 0},clip,width=0.5\linewidth]
{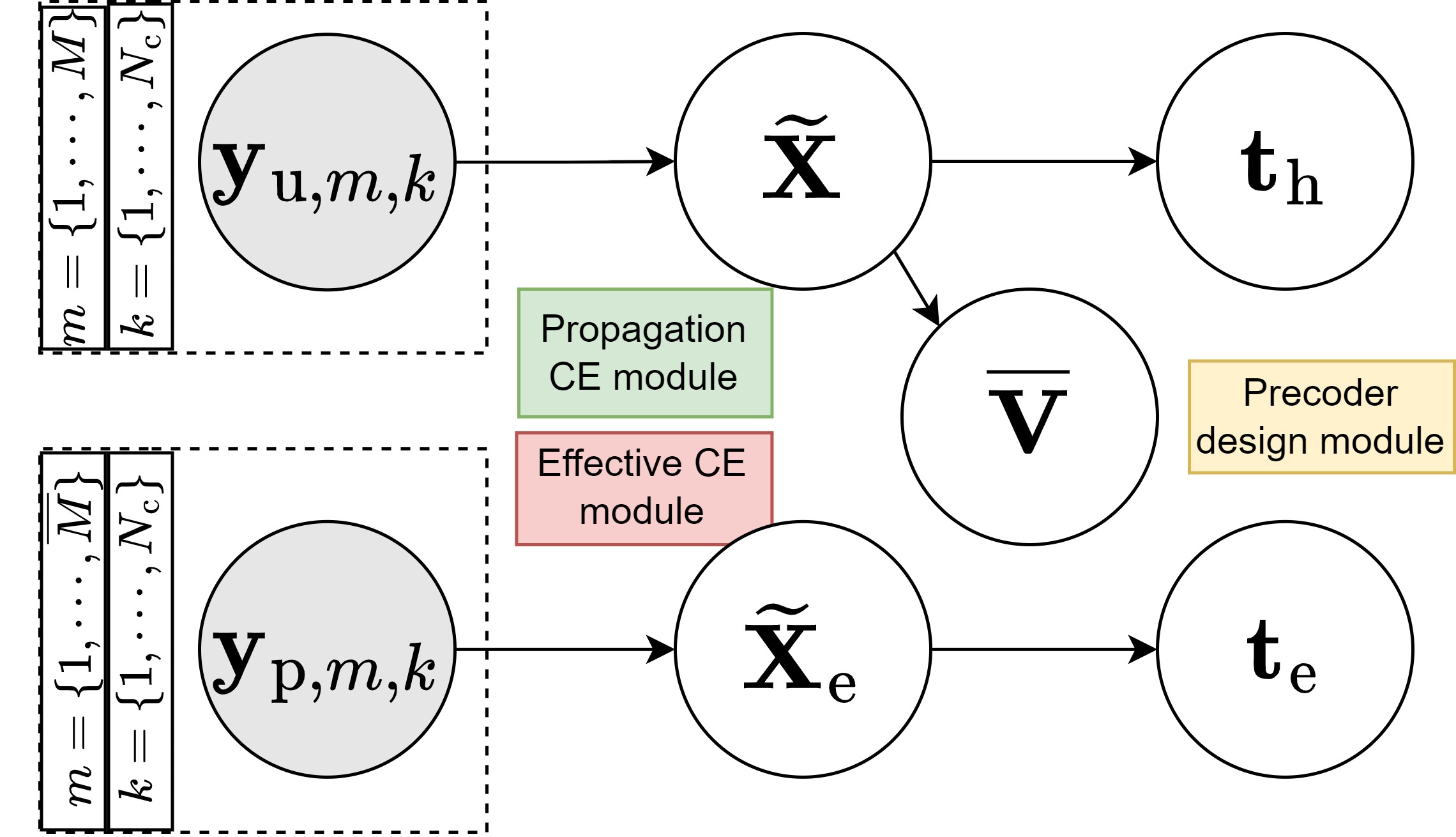}
		}
    \caption{(a.) Probabilistic graphical model for the generative model which captures the observations corresponding to the labelled data, pilot observations at the transmitter and at the receiver, as well as the Markov prior for the channel model.  (b.) Inference model which captures the functions of various modules in the datapath.}\label{fig::prior-model}
\end{figure}

\subsubsection{Likelihood distributions}
The likelihood distributions consist of the pilot observations and the labels for the payloads:
\begin{equation}\label{eq::lik}
\begin{aligned}[b]
&\p_\mathrm{lik}(\mathbfcal{Y}_\mathrm{u}, \mathbfcal{Y}_\mathrm{p}, \boldsymbol{\beta}| \mathbf{\widetilde{X}}, \mathbf{\widetilde{X}}_\mathrm{e},\mathbfcal{Y}_{\mathrm{d}})=\p(\mathbfcal{Y}_\mathrm{u}| \mathbf{\widetilde{X}})
 \p(\mathbfcal{Y}_\mathrm{p}|\mathbf{\widetilde{X}}_\mathrm{e})
\p(\boldsymbol{\beta} |\mathbfcal{Y}_{\mathrm{d}}, \mathbf{\widetilde{X}}_{\mathrm{e}}).
\end{aligned}
\end{equation}

\begin{itemize}
\item \textbf{SRS observations} $\p(\mathbfcal{Y}_\mathrm{u}| \mathbf{\widetilde{X}})$: The likelihood for SRS is obtained from \eqref{eq::feed_start_trunc}:
\begin{align}
\p(\mathbfcal{Y}_\mathrm{u}| \mathbf{\widetilde{X}} ) &= \prod_{m=1}^{M} \prod_{k=1}^{K}  \p(\mathbf{y}_{\mathrm{u},m, k}|\mathbf{\widetilde{X}} ),\\
\p(\mathbf{y}_{\mathrm{u}, m, k}| \mathbf{\widetilde{X}} ) &=  \mathcal{CN}( \mathbf{H}_k^\mathrm{T} \sum_{n_{\rm r}} \alpha_{m, k, n_{\rm r}} {\bf s}_{{\rm p},n_{\rm r}}, \sigma_\mathrm{u}^2 \mathbf{I}_{N_\mathrm{t}}),\\
\mathbf{H}_k &=  ( \mathbf{e}_{K}^{(k)} \otimes \mathbf{I}_{N_\mathrm{r}}) (\widetilde{\mathbf{F}}_D  \otimes \mathbf{I}_{N_\mathrm{r}}) \mathbf{\widetilde{X}}.
\end{align}

\item  \textbf{DMRS pilot observations} $\p(\mathbfcal{Y}_\mathrm{p}| \mathbf{\widetilde{X}}_\mathrm{e})$: Similarly, likelihood for observations corresponding to DMRS is given by:
\begin{align}
\p(\mathbfcal{Y}_\mathrm{p}| \mathbf{\widetilde{X}}_\mathrm{e} )&=  \prod_{\overline{m}=1}^{\overline{M}} \prod_{k=1}^{K}  \p(\mathbf{y}_{\mathrm{p},m, k}|\mathbf{\widetilde{X}}_\mathrm{e}),\\
\p(\mathbf{y}_{\mathrm{p},\overline{m},k}| \mathbf{\widetilde{X}}_\mathrm{e} ) &=  \mathcal{CN}(\mathbf{H}_{\mathrm{e}, k}^\mathrm{T} \sum_\ell \overline{\alpha}_{\overline{m}, k, \ell} {\bf \overline{s}}_{{\rm p},\ell}, \sigma_\mathrm{r}^2 \mathbf{I}_{N_\mathrm{r}}),\\
\mathbf{H}_{\mathrm{e},k} &=  ( \mathbf{e}_{K}^{(k)} \otimes \mathbf{I}_{N_\mathrm{r}}) (\widetilde{\mathbf{F}}_{D_\mathrm{eff}}  \otimes \mathbf{I}_{N_\mathrm{r}}) \mathbf{\widetilde{X}}_\mathrm{e},
\end{align}
 where $\widetilde{\mathbf{F}}_{D_\mathrm{eff}} \in \mathbb{C}^{K \times {D_\mathrm{eff}}}$ is the partial DFT matrix,  created by truncating $\mathbf F$ to keep only the first $D_\mathrm{eff}$ columns.

%
%

\item \textbf{Payload Demodulation} $\p(\boldsymbol{\beta} |\mathbfcal{Y}_{\mathrm{d}}, \mathbf{\widetilde{X}}_{\mathrm{e}})$:
We then model the labelled data distribution as:
\begin{equation}
\begin{aligned}\label{eq::datadet}
\p(\boldsymbol{\beta} |\mathbfcal{Y}_{\mathrm{d}}, \mathbf{\widetilde{X}}_{\mathrm{e}})=
\mathbb{E}_{\boldsymbol{\rho}}\left[\p(\boldsymbol{\beta} | \boldsymbol{\rho}) \p(\boldsymbol{\rho}|\mathbfcal{Y}_{\mathrm{d}}, \mathbf{\widetilde{X}}_{\mathrm{e}})\right],
\end{aligned}
\end{equation}
where:
\begin{align}
\p(\boldsymbol{\beta} | \boldsymbol{\rho}) = \prod_{m=1}^{M}\prod_{k=1}^{K}\prod_{\ell=1}^{L}\prod_{b=1}^{B} \psi_\mathrm{Brl}({\beta}_{m,k,\ell,b}; \rho_{m,k,\ell,b}),
\end{align}
where $\psi_\mathrm{Brl}$ is the Bernoulli distribution. Specifically, from the data observations in \eqref{eq::sys_ydata} and \eqref{eq::sys_rk}, the estimated data bit probabilities is given by \eqref{eq::demodulate}:
\begin{equation}\label{eq::demodlearn}
\p(\boldsymbol{\rho}|\mathbfcal{Y}_{\mathrm{d}}, \mathbf{\widetilde{X}}_{\mathrm{e}}) = \prod_{m=1}^{M}\prod_{k=1}^{K}\prod_{\ell=1}^L \delta({\bm \rho}_{m,k,\ell} - \mathrm{g}^{-1}_{\mathrm{q},B}(\hat{r}_{m,k,\ell})).
\end{equation}


\end{itemize}

\subsubsection{Prior distributions}

The prior distribution acts as a regularization term, ensuring that the key variables display specific characteristics aligned with domain knowledge:
\begin{equation}\label{eq::pri}
\begin{aligned}[b]
&\p_\mathrm{pri}\bigl(\mathbf{t}_\mathrm{h}, \mathbf{\widetilde{X}},\mathbf{\overline{V}}, \mathbf{t}_\mathrm{e}, \mathbf{\widetilde{X}}_\mathrm{e};{\bm \omega}_\mathrm{pir} \bigr)
=\p(\mathbf{t}_\mathrm{h}; {\bm \omega}_\mathrm{shp})\\
&\p(\mathbf{\widetilde{X}}|\mathbf{t}_\mathrm{h}; {\bm \omega}_\mathrm{shv})
\p(\mathbf{\overline{V}}|\mathbf{\widetilde{X}})\p(\mathbf{t}_\mathrm{e}; {\bm \omega}_\mathrm{sep})\p(\mathbf{\widetilde{X}}_\mathrm{e}|\mathbf{t}_\mathrm{e}; {\bm \omega}_\mathrm{sev}),
\end{aligned}
\end{equation}
where ${\bm \omega}_\mathrm{pir} = \{{\bm \omega}_\mathrm{shp}, {\bm \omega}_\mathrm{shv}, {\bm \omega}_\mathrm{sep}, {\bm \omega}_\mathrm{sev}\}$.

\begin{itemize}
\item  \textbf{Delay-Domain CSIT} $\p(\mathbf{\widetilde{X}}|\mathbf{t}_\mathrm{h}; {\bm \omega}_\mathrm{shv})$: The propagation channel is modelled by:
\begin{equation}\label{eq::pri_csit}
\p(\mathbf{\widetilde{X}}|\mathbf{t}_\mathrm{h}; {\bm \omega}_\mathrm{shv}) = \prod_{d=1}^{D} \psi_\mathrm{mg}(\mathbf{{X}}_d; {t}_{\mathrm{h},d}, \epsilon_{\rm{h0}}, {\epsilon}_{{\rm{h1}},d}),
\end{equation}
where $\mathbf{{X}}_d = (\mathbf{e}_{D}^{(d)} \otimes \mathbf{I}_{N_\mathrm{r}}) \mathbf{\widetilde{X}}$ is the delay domain  channel at the $d$-th delay tap, and $\psi_\mathrm{mg}(\cdot)$ denotes the mixture Gaussian distribution:
\begin{equation}\begin{aligned}[b]
\psi_\mathrm{mg}(\mathbf{{X}}; t, \epsilon_{0}, \epsilon_{1})&=(1-t) \mathcal{CMN} ({\bm 0};\epsilon_{0}\mathbf{I}_{N_1}, \epsilon_{0}\mathbf{I}_{N_2}) \\
&+ t \mathcal{CMN} ({\bm 0}; \epsilon_{1}\mathbf{I}_{N_1}, \epsilon_{1}\mathbf{I}_{N_2}),
\end{aligned}
\end{equation}
with any $\mathbf{X} \in {\mathbb C}^{N_1 \times N_2}$ and $t \in \{0,1\}$ denotes the support.
The parameters $\{\epsilon_{h0}, \boldsymbol{\epsilon}_{h1}\}$ in \eqref{eq::pri_csit} are learnt using a propagation variance supplementary generative network illustrated in \Figref{fig::supp:prop}{(a)} with weights ${\bm \omega}_\mathrm{shv}$:
\begin{align}
\{\epsilon_{h0}, \boldsymbol{\epsilon}_{h1}\}  &= \mathrm{g}_\mathrm{sv}(\mathbfcal{Y}_\mathrm{u}, \sigma_\mathrm{u}^2, D; {\bm \omega}_\mathrm{shv}).
\end{align}

\begin{figure}[!t]
    \centering
    \subfigure[Variance supplementary generative network]    {
	    	\includegraphics[trim={0 0 0 0},clip,width=0.45\linewidth]{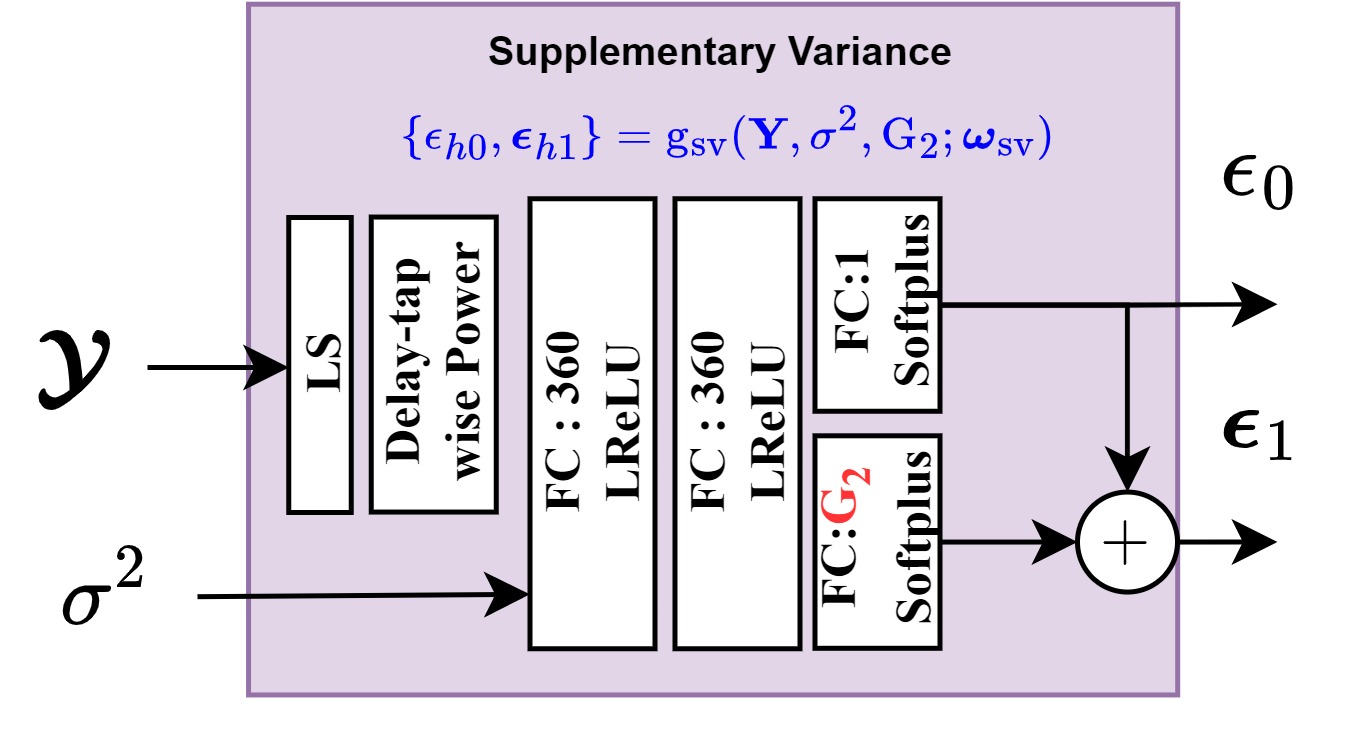}
		}
    \subfigure[Support supplementary generative network]    {
	    	\includegraphics[trim={0 0 0 0},clip,width=0.45\linewidth]{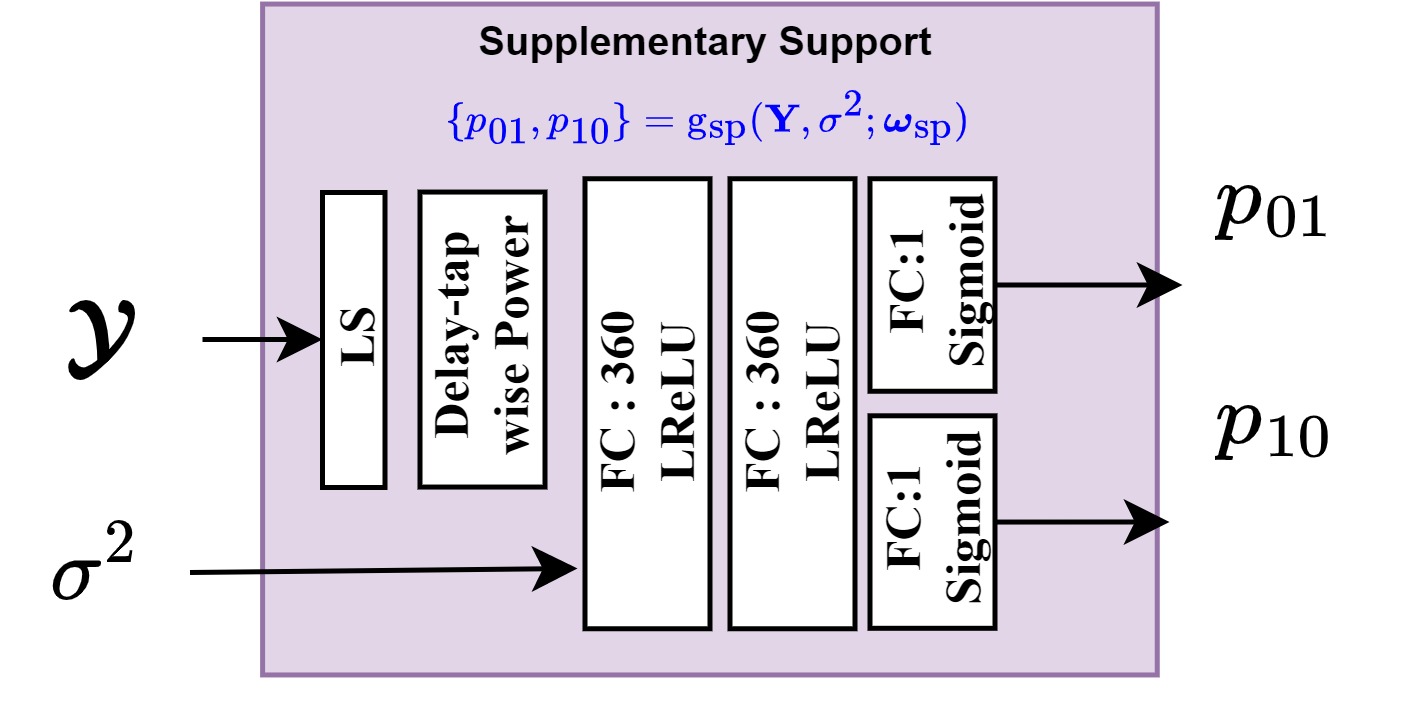}
	    	}
	
	\caption{ Supplementary networks parameterising the generative distribution parameters (a.) Variance parameters for channel, (b.) Markov distribution for channel support.  In traditional VBI for learning the variance parameters, one can initialize $\boldsymbol{\epsilon}_{1, k} > \sigma_{0}, \forall k \in \{1,...,\mathrm{G}_2\}$ to ensure a good convergence. For the trainable networks, we enforce it by making $\sigma_{1}$ as a sum of $\sigma_{0}$ and an intermediate positive value.}\label{fig::supp:prop}
\end{figure}

\item \textbf{Delay-Domain CSIT Support} $\p(\mathbf{t}_\mathrm{h}; {\bm \omega}_\mathrm{shp})$: To model the clustered sparsity in the delay domain, we have:
\begin{equation}\label{eq::pri_csit_supp}
\p(\mathbf{t}_\mathrm{h};{\bm \omega}_\mathrm{shp}) = \psi_\mathrm{Mar}(\mathbf{t}_\mathrm{h}; D, p_{h01}, p_{h10}),
\end{equation}
where $\psi_\mathrm{Mar}$ is the Markov distribution as:
\begin{align}
&\psi_{\mathrm{Mar}}(\mathbf{t}; K_{\rm T}, p_{01}, p_{10}) =\p(t_\mathrm{1}) \prod_{k=2}^{K_{\rm T}}  \p(t_\mathrm{k}|t_\mathrm{k-1}),\\
&\p(t_\mathrm{1}) = p_\mathrm{avg}^{t_\mathrm{1}}(1-p_\mathrm{avg})^{1-t_\mathrm{1}},\\
&\p(t_\mathrm{k}| t_\mathrm{k-1}) =
\begin{cases}
(1-p_{01})^{1-t_\mathrm{d}} p_{01}^{t_\mathrm{d}}, & {\rm if}\; t_\mathrm{d-1}=0,\\
p_{h10}^{1-t_\mathrm{d}} (1-p_{h10})^{t_\mathrm{d}}, &{\rm if}\; t_\mathrm{d-1}=1,
\end{cases}
\end{align}
with $p_\mathrm{avg}=\frac{p_{01}}{p_{01}+p_{10}}$.
The parameters $p_{h01}$ and  $p_{h10}$ in \eqref{eq::pri_csit_supp} are learnt using a propagation support supplementary generative network illustrated in \Figref{fig::supp:prop}{(b)} with weights ${\bm \omega}_\mathrm{shp}$:
\begin{equation}
\{p_{h01}, p_{h10}\} = \mathrm{g}_\mathrm{sp}(\mathbfcal{Y}_\mathrm{u}, \sigma_\mathrm{u}^2; {\bm \omega}_\mathrm{shp}),
\end{equation}
where the architecture of $\mathrm{g}_\mathrm{shp}$ is illustrated in \Figref{fig::supp:prop}{a}.

\item  \textbf{Precoder} $\p(\mathbf{\overline{V}}|\mathbf{\widetilde{X}})$:
The precoder distribution is composed using the power constriant in the delay domain, followed by the frequency domain transform as:
\begin{align}
\p(\mathbf{\overline{V}}|\mathbf{\widetilde{X}})&
=  \mathbb{E}_{\mathbf{\widetilde{W}}} \left[ \p(\mathbf{\overline{V}}|\mathbf{\widetilde{W}}) \p(\mathbf{\widetilde{W}}|\mathbf{\widetilde{X}})\right],\\
\p(\mathbf{\widetilde{W}}|\mathbf{\widetilde{X}}) &= \delta(\mathbf{\widetilde{W}} - \frac{\mathbf{\widetilde{W}}}{\|\mathbf{\widetilde{W}}\|_\mathrm{F}} \sqrt{K P_\mathrm{T}}),\\
\p(\mathbf{\overline{V}}|\mathbf{\widetilde{W}}) &= \delta(\mathbf{\overline{V}} - (\widetilde{\mathbf{F}}_\mathrm{\overline{v}} \otimes \mathbf{I}_{N_\mathrm{t}}) \mathbf{\widetilde{W}}),
\end{align}
where $\widetilde{\mathbf{F}}_\mathrm{\overline{v}} \in \mathbb{C}^{K \times {D}_{\rm v}}$ is the partial DFT matrix by truncating $\mathbf F$ to retain only the first ${D}_{\rm v}$ columns, and ${D}_\mathrm{v}=\|\mathbf{\overline{e}}_{D_\mathrm{eff}}^{({D}_{\mathrm{v}})}\|_0$ is defined based on \eqref{eq::prec:supp_e}.

\item \textbf{Delay-domain CSIR} $\p(\mathbf{\widetilde{X}_\mathrm{e}}|\mathbf{t}_\mathrm{e}; {\bm \omega}_\mathrm{sev})$: As before, the effective channel distribution is:
\begin{equation}
\p(\mathbf{\widetilde{X}}_\mathrm{e}|\mathbf{t}_\mathrm{e}; {\bm \omega}_\mathrm{sev}) = \prod_{d=1}^{{D_\mathrm{eff}}} \psi_\mathrm{mg}(\mathbf{{X}}_{\mathrm{e},d}; {t}_{\mathrm{e},d}, \epsilon_{e0}, \boldsymbol{\epsilon}_{e1}),
\end{equation}
where $\mathbf{{X}}_{\mathrm{e},d} = (\mathbf{e}_{{D_\mathrm{eff}}}^{(d)} \otimes \mathbf{I}_{N_\mathrm{r}}) \mathbf{\widetilde{X}}_{\mathrm{e}}$ is the delay domain effective channel at the $d$-th delay tap with parameters learnt using a propagation variance supplementary generative network with weights ${\bm \omega}_\mathrm{sev}$:
\begin{align}
\{\epsilon_{e0}, \boldsymbol{\epsilon}_{e1} \} &= \mathrm{g}_\mathrm{sv}(\mathbfcal{Y}_\mathrm{p}, \sigma_\mathrm{r}^2, {D_\mathrm{eff}}; {\bm \omega}_\mathrm{sev}).
\end{align}

\item  \textbf{Delay-domain CSIR Support} $\p(\mathbf{t}_\mathrm{e}; \mathbf{w}_\mathrm{sep})$: The precoder design enforces a sparse effective channel. Since the propagation channel is clustered and the precoder bounded, we expect the effective channel to be clustered and we model it using the Markov distribution:
\begin{equation}
\p(\mathbf{t}_\mathrm{e}; {\bm \omega}_\mathrm{shp}) = \psi_\mathrm{Mar}(\mathbf{t}_\mathrm{e}; {D_\mathrm{eff}}, p_{e01}, p_{e10}),
\end{equation}
where the parameters of the Markov distribution are learnt using an effective support supplementary generative network with weights ${\bm \omega}_\mathrm{sep}$:
\begin{align}
\{p_{e01}, p_{e10}\}  = \mathrm{g}_\mathrm{sp}(\mathbfcal{Y}_\mathrm{p}, \sigma_\mathrm{r}^2; {\bm \omega}_\mathrm{sep}).
\end{align}

\end{itemize}

\subsection{The Inference Model}\label{sec::inf}
The inference model encompasses all three designed DNN modules:
\begin{equation}\label{eq::inf}
\begin{aligned}[b]
\q(\mathbf{t}_\mathrm{h},\mathbf{\widetilde{X}},  \overline{\mathbf{V}}, &\mathbf{t}_\mathrm{e}, \mathbf{\widetilde{X}}_\mathrm{e}; {\bm \omega}_\mathrm{inf}) = \q(\mathbf{\widetilde{X}}; {\bm \omega}_\mathrm{h}^{(1)})\q(\mathbf{t}_\mathrm{h}|\mathbf{\widetilde{X}}; {\bm \omega}_\mathrm{h}^{(2)})\\&\q(\overline{\mathbf{V}}; {\bm \omega}_{\mathrm{v},L}) \q(\mathbf{\widetilde{X}}_\mathrm{e}; {\bm \omega}_\mathrm{e}^{(1)})\q(\mathbf{t}_\mathrm{e}|\mathbf{\widetilde{X}}_\mathrm{e}; {\bm \omega}_\mathrm{e}^{(2)}),
\end{aligned}
\end{equation}
and we further define  ${\bm \omega}_\mathrm{inf} = \{{\bm \omega}_\mathrm{h}, {\bm \omega}_{\mathrm{v},L},{\bm \omega}_\mathrm{e}\}$ for all the weights in the proposed DNN datapath.

\begin{itemize}
\item  \textbf{Propagation Channel Estimation} $\q(\mathbf{\widetilde{X}}; {\bm \omega}_\mathrm{h}^{(1)})$: The estimated posterior distribution of the CSIT is modelled as an element-wise matrix complex normal distribution. Recall $\boldsymbol{\mu}_{\mathbf{\widetilde{X}}}$ and $\boldsymbol{\sigma}_{\mathbf{\widetilde{X}}}$ from \eqref{eq::datapath_prop_mu_sigma}, yielding:
\begin{equation}
\q(\mathrm{vec}(\mathbf{\widetilde{X}}); {\bm \omega}_\mathrm{h}) = \mathcal{CN}(\mathrm{vec}(\boldsymbol{\mu}_{\mathbf{\widetilde{X}}}), \mathrm{diag}(\mathrm{vec}(\boldsymbol{\sigma}_{\mathbf{\widetilde{X}}})^2)).
\end{equation}

\item \textbf{Propagation Channel Support} $\q(\mathbf{t}_\mathrm{h}|\mathbf{\widetilde{X}}; {\bm \omega}_\mathrm{h}^{(2)})$: The channel support is modelled by Bernoulli distribution:
\begin{equation}
\q(\mathbf{t}_\mathrm{h}|\mathbf{\overline{X}}; {\bm \omega}_\mathrm{h}) = \prod_{d=1}^{D} \psi_\mathrm{Brl}(t_{\mathrm{h},d}; \lambda_{\mathrm{h},d}),
\end{equation}
where $\boldsymbol{\lambda}_\mathrm{h}$ is a function of  $\mathbf{\widetilde{X}}$ as shown in \eqref{eq::datapath_prop_lam}.

\item \textbf{Precoder} $\q(\overline{\mathbf{V}}|\mathbf{\widetilde{X}}; {\bm \omega}_{\mathrm{v},L})$: The precoder is defined as a deterministic function. Recall that $\mathbf{\widetilde{W}}$ in \eqref{eq::precoder_final} is a function of  $\mathbf{\widetilde{X}}$, we have:
\begin{equation}
\q(\overline{\mathbf{V}}|\mathbf{\widetilde{X}}; {\bm \omega}_{\mathrm{v},L}) = \delta(\overline{\mathbf{V}} - (\widetilde{\mathbf{F}}_\mathrm{\overline{v}} \otimes \mathbf{I}_{N_\mathrm{t}})\mathrm{g}_\mathrm{v}\mathbf{\widetilde{W}}).
\end{equation}

\item  \textbf{Effective Channel Estimation} $\q(\mathbf{\widetilde{X}}_\mathrm{e};{\bm \omega}_\mathrm{e}^{(1)})$: Similar to the propagation channel, the posterior distribution is modelled as an element-wise matrix complex normal distribution. Using $\boldsymbol{\mu}_{\mathbf{\widetilde{X}_\mathrm{e}}}$ and $\boldsymbol{\sigma}_{\mathbf{\widetilde{X}_\mathrm{e}}}$ from \eqref{eq::datapath_eff_mu_sigma}, we have:
\begin{equation}
\q(\mathrm{vec}(\mathbf{\widetilde{X}}_\mathrm{e}); {\bm \omega}_\mathrm{e}) = \mathcal{CN}(\mathrm{vec}(\boldsymbol{\mu}_{\mathbf{\widetilde{X}}_\mathrm{e}}), \mathrm{diag}(\mathrm{vec}(\boldsymbol{\sigma}_{\mathbf{\widetilde{X}}_\mathrm{e}})^2)).
\end{equation}

\item  \textbf{Effective Channel Support} $\q(\mathbf{t}_\mathrm{e}|\mathbf{\widetilde{X}}_\mathrm{e}; {\bm \omega}_\mathrm{e}^{(2)})$: The channel support is modelled by Bernoulli distribution as:
\begin{equation}
\q(\mathbf{t}_\mathrm{e}|\mathbf{\widetilde{X}}_\mathrm{e}; {\bm \omega}_\mathrm{v}) = \prod_{d=1}^{{D_\mathrm{eff}}} \psi_\mathrm{Brl}(t_{\mathrm{e},d}; \lambda_{\mathrm{e},d}),
\end{equation}
where $\boldsymbol{\lambda}_\mathrm{e}$ is obtained from \eqref{eq::datapath_eff_lam} as a function of $\mathbf{\widetilde{X}}_\mathrm{e}$.

\end{itemize}

\subsection{Training Algorithm}
\subsubsection{Formulation of the VBI Loss Function}
Given the generative and inference models mentioned above, we design the end-to-end objective by maximizing the evidence lower bound (ELO), which is equivalent to minimizing the KL divergence. Consider a dataset of $J$ samples of labelled data bits and received observation symbols indexed by $j$:
\begin{equation}
\mathcal{D} = \{ \mathbfcal{Y}_\mathrm{u}^{(j)}, \mathbfcal{Y}_\mathrm{p}^{(j)}, \mathbfcal{Y}_\mathrm{d}^{(j)},  \boldsymbol{\beta}^{(j)} \}.
\end{equation}
Then for the $j$-th sample $\mathcal{D}^{(j)}$, the ELO objective is given by:
\begin{equation}\label{eq::elbo_loss}
\begin{aligned}[b]
    &\mathrm{f}_\mathrm{ELO}(  {\bm \omega}, \mathcal{D}^{(j)}) = \mathbb{E}_{\q(\mathbf{t}_\mathrm{h}^{(j)},\mathbf{\widetilde{X}}^{(j)},  \overline{\mathbf{V}}^{(j)}, \mathbf{t}_\mathrm{e}^{(j)}, \mathbf{\widetilde{X}}_\mathrm{e}^{(j)}; {\bm \omega}_\mathrm{inf})}  \mathrm{log} \Bigl[\\
    &\dfrac{
        \makecell{
            \p\bigl(\mathbf{t}_\mathrm{h}^{(j)}, \mathbf{\widetilde{X}}^{(j)}, \mathbfcal{Y}_\mathrm{u}^{(j)}, \mathbf{\widetilde{W}}^{(j)}, \mathbf{t}_\mathrm{e}^{(j)}, \mathbf{\widetilde{X}}_\mathrm{e}^{(j)}, \mathbfcal{Y}_\mathrm{p}^{(j)}, \boldsymbol{\beta}^{(j)}| \mathbfcal{Y}_\mathrm{d}^{(j)} ; {\bm \omega}_\mathrm{pir}\bigr)
    }
    }{
    \q(\mathbf{t}_\mathrm{h}^{(j)},\mathbf{\widetilde{X}}^{(j)},  \overline{\mathbf{V}}, \mathbf{t}_\mathrm{e}^{(j)}, \mathbf{\widetilde{X}}_\mathrm{e}^{(j)}; {\bm \omega}_\mathrm{inf})
    }\Bigr],
    \end{aligned}\end{equation}
    where the detailed distributions have been designed in \eqref{eq::gen} and \eqref{eq::inf}, respectively, and ${\bm \omega} = \{{\bm \omega}_\mathrm{pir}, {\bm \omega}_\mathrm{inf}\}$. The expectation operation in \eqref{eq::elbo_loss} is implemented numerically through sampling \cite{kingma2013auto} and reparameterization tricks \cite{jang2016categorical}.
    The overall objective is then given by:
    \begin{subequations}\label{eq::elo_final}
    \begin{align}
    \nonumber
    {\mathcal{P}}{(\text{B})}:\,\, & \underset{{\bm \omega}}{\mathrm{max}}  \;  \;  \sum_{j=1}^{J} \mathrm{f}_\mathrm{ELO}(  {\bm \omega}, \mathcal{D}^{(j)}).
    \end{align}
    \end{subequations}

\subsubsection{Training Process Details}
We implement the DNN training using Tensorflow \cite{tensorflow2015}, generating $10^5$ random channel realizations for training and $10^4$ for testing.  We set $D_\mathrm{eff}=\frac{K}{8}$ to obtain a sparse effective channel, and the number of unrolling iterations is set to ${\overline{T}}=3$.
Since the end-to-end design consists of multiple components, pre-training the framework ensures a solid initialization. We conduct pre-training in three steps:
\begin{enumerate}
\item First, we train the precoder design module to minimize the BER with perfect CSIT and CSIR, while applying a hard delay-domain truncation threshold of $D_\mathrm{v}=56$.
\item Next, we separately train the propagation channel estimation and effective channel estimation modules based on the NMSE metric, keeping the precoder fixed.
\item Finally, we fix the weights of the first three modules and train the supplementary networks using the ELO loss function in \eqref{eq::elbo_loss}.
\end{enumerate}

After pre-training, we proceed with end-to-end training to address ${\mathcal{P}}{(\text{B})}$. The loss is optimized using the Adam optimizer \cite{kingma2014adam} with a learning rate of $10^{-3}$, which decays over 50 epochs. Subsequently, we fine-tune the model using stochastic gradient descent with an initial learning rate of $10^{-5}$ for 150 epochs, also applying learning rate decay.


\section{Simulation Results}
\subsection{Setups}
We consider the 802.11ax MIMO channel with the delay profile configured to model B, using a carrier frequency of 2.4 GHz. The sub-carrier spacing is set to 30 KHz, with $K=1024$ sub-carriers and a total bandwidth of 80 MHz. We evaluate the performance for two antenna setups: setup A with $N_\mathrm{t}=8$ and $N_\mathrm{r}=8$, and setup B with $N_\mathrm{t}=16$ and $N_\mathrm{r}=16$.  We observe a maximum delay spread of $D=72$, leading to a SRS scheme wtih $A_\mathrm{u}=8$ for CSIT estimation, and we set $A_\mathrm{r}=8$ for the DMRS proposed in \eqref{eq::sparse_dmrs}.

\begin{table}[!t]
\footnotesize
\renewcommand{\arraystretch}{1.1}
\caption{\textcolor{black}{Detailed complexity of different modules of the proposed NN based design for $N_\mathrm{r}=8, N_\mathrm{r}=8, L=4$}}
\label{tab::complexity}
\centering
\begin{tabular}{|l|l|l|}
\hline
\textbf{Module} & \textbf{Component} & \textbf{FLOP} \\ \hline\hline
\multirow{2}{*}{Propagation CE Module} &   Channel Estimation   &   57.7M \\ \cline{2-3}
                                       &   Active Delay   &    181K \\ \hline
\multirow{4}{*}{Precoder Design Module} &   Lagrange Network &  125K   \\ \cline{2-3}
                              &   Adaptive Window   &   1.6M   \\ \cline{2-3}
                              &   Unrolling   &  262.5M  \\ \cline{2-3}
                              &   NN based Initializer &  65.2M  \\ \hline
\multirow{2}{*}{Effective CE Module} &   Channel Estimation   &  50.4M  \\ \cline{2-3}
                                 &   Active Delay   &   321K  \\ \hline
\end{tabular}
\end{table}

\begin{table}[!t]
\footnotesize
\renewcommand{\arraystretch}{1.1}
\caption{\textcolor{black}{Detailed complexity of different modules of the proposed DNN based design for $N_\mathrm{r}=8, N_\mathrm{r}=8, L=4$}}
\label{tab::complexity_time}
\centering
\begin{tabular}{|l||l|}
\hline
\textbf{Baseline} & \textbf{Time (ms)}  \\ \hline\hline
Common + VAMP    &   23.5 				\\ \hline
S-WMMSE + VAMP     &      817.2		    \\ \hline
Proposed EVM + VAMP           &      814.3	    	    \\ \hline
Proposed	DNN	     &      17.1	    		    \\ \hline
\end{tabular}
\end{table}

\subsection{Baselines}
For the channel estimation evaluation, we assess performance against the following baselines:
\begin{enumerate}
\item {\textbf{Least Squares (LS)} }: We note that the number of pilot observations matches the number of channel variables, allowing for least squares estimation.
\item {\textbf{Orthogonal Matching Pursuit (OMP)}} \cite{tropp2007signal}: Orthogonal matching pursuit is a fast algorithm for sparse estimation.
\item {\textbf{Vector Approximate Message Passing (VAMP)}} \cite{wang2022channel}: This baseline uses VAMP as a fast iterative sparse approximation.
\item {\textbf{LASSO}} \cite{ranstam2018lasso}:  The Lasso is a popular sparse estimation baseline which uses an $\ell_1$ based regularisation.
\item {\textbf{Genie-aided MMSE (GA-MMSE)}}: This baseline uses the true channel variance as the performance benchmark.
\end{enumerate}

For the end-to-end performance evaluation, we fix the uplink SNR at 20 dB. Since sparse precoding is crucial for achieving a sparse effective channel, we consider the following sparse precoding baselines:
\begin{enumerate}
\item {\textbf{Common Precoder}}: A common precoder is optimized for all sub-carriers as introduced in Section \ref{sec::comm_sparse_precoder}.
\item {\textbf{Sparse WMMSE (S-WMMSE)}} \cite{zhang2023cross}: The delay-domain precoder is optimized to maximize the same rate across all sub-carriers with $D_\mathrm{v} = 56$. We utilize the best sparse recovery baseline algorithm for estimating CSIT and CSIR.
\end{enumerate}
Additionally, we will present the performance of two proposed solutions:
\begin{enumerate}
\item {\textbf{Proposed EVM}}: The delay-domain precoder is optimized by minimizing the EVM by solving ${\mathcal{P}}{(\text{A})}$ with $D_\mathrm{v} = 56$. We utilize the best sparse recovery baseline algorithm for estimating CSIT and CSIR.
\item {\textbf{Proposed DNN}}: The end-to-end DNN solution described in Section \ref{sec:detapath} is trained by solving $\mathcal{P}(\text{B})$.
\end{enumerate}


The complexity of the various NN modules is summarized in Table \ref{tab::complexity} and \ref{tab::complexity_time}. The overall FLOP required is around 0.4G FLOPs which requires a total of around 17 ms for CPU.


 \begin{figure}[!t]
    \centering
	    	\includegraphics[trim={0 0 0 0},clip,width=0.8\linewidth]{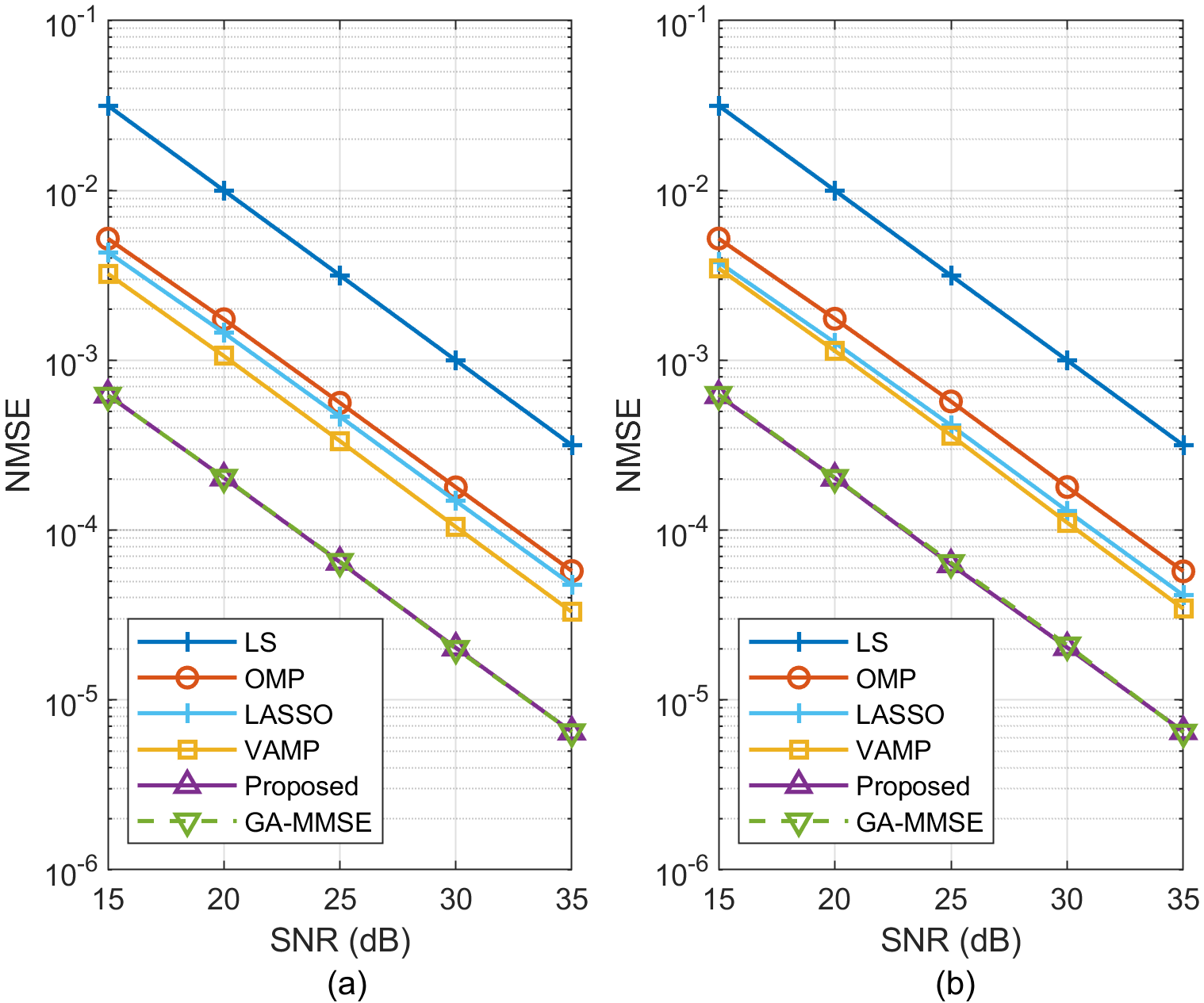}
	\caption{NMSE for CSIT estimation: (a) setup A; (b) setup B. }\label{fig::res_csit}
\end{figure}

 \begin{figure}[!t]
    \centering
	    	\includegraphics[trim={0 0 0 0},clip,width=0.9\linewidth]{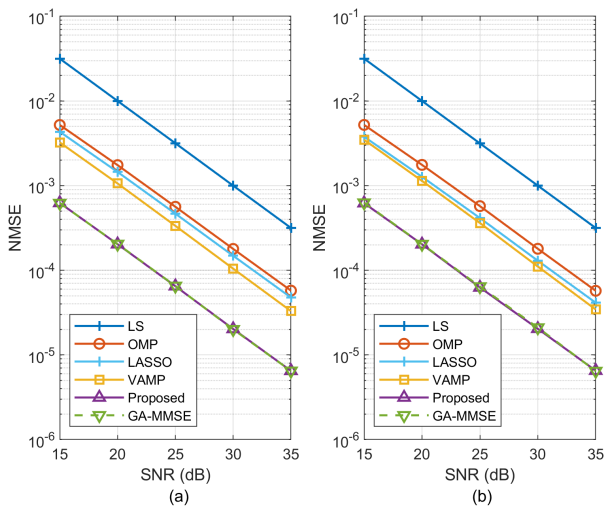}
	\caption{NMSE for CSIR estimation with DNN-based transmitter setting $L = N_{\rm t}/2$: (a) setup A; (b) setup B.}\label{fig::res_csir}
\end{figure}

\subsection{Simulation Results}

\subsubsection{Channel Estimation}
 \Figref{fig::res_csit}{} and \Figref{fig::res_csir}{} compare the NMSE of the estimated CSIT and CSIR both the baselines and our proposed solution. Our approach closely matches the performance of the Genie-aided MMSE, surpassing the LS baselines by over 15 dB and exceeding all sparse channel recovery baselines by more than 7 dB. This demonstrates that our solution provides an interpretable data path design, which existing end-to-end DNN designs \cite{sohrabi2021deep, jang2019deep,  wu2022deep, zhang2022deep, attiah2022deep, pellaco2020deep,jang2022deep, shi2023robust, jin2023model,hu2021two} cannot guarantee.
 Additionally, VAMP performs best among the baselines, so we adopt it for the common precoder and S-WMMSE in the end-to-end simulation.


\subsubsection{BER for Different Number of Streams}
\begin{figure}[!t]
    \centering
	    	\includegraphics[trim={0 0 0 0},clip,width=0.8\linewidth]{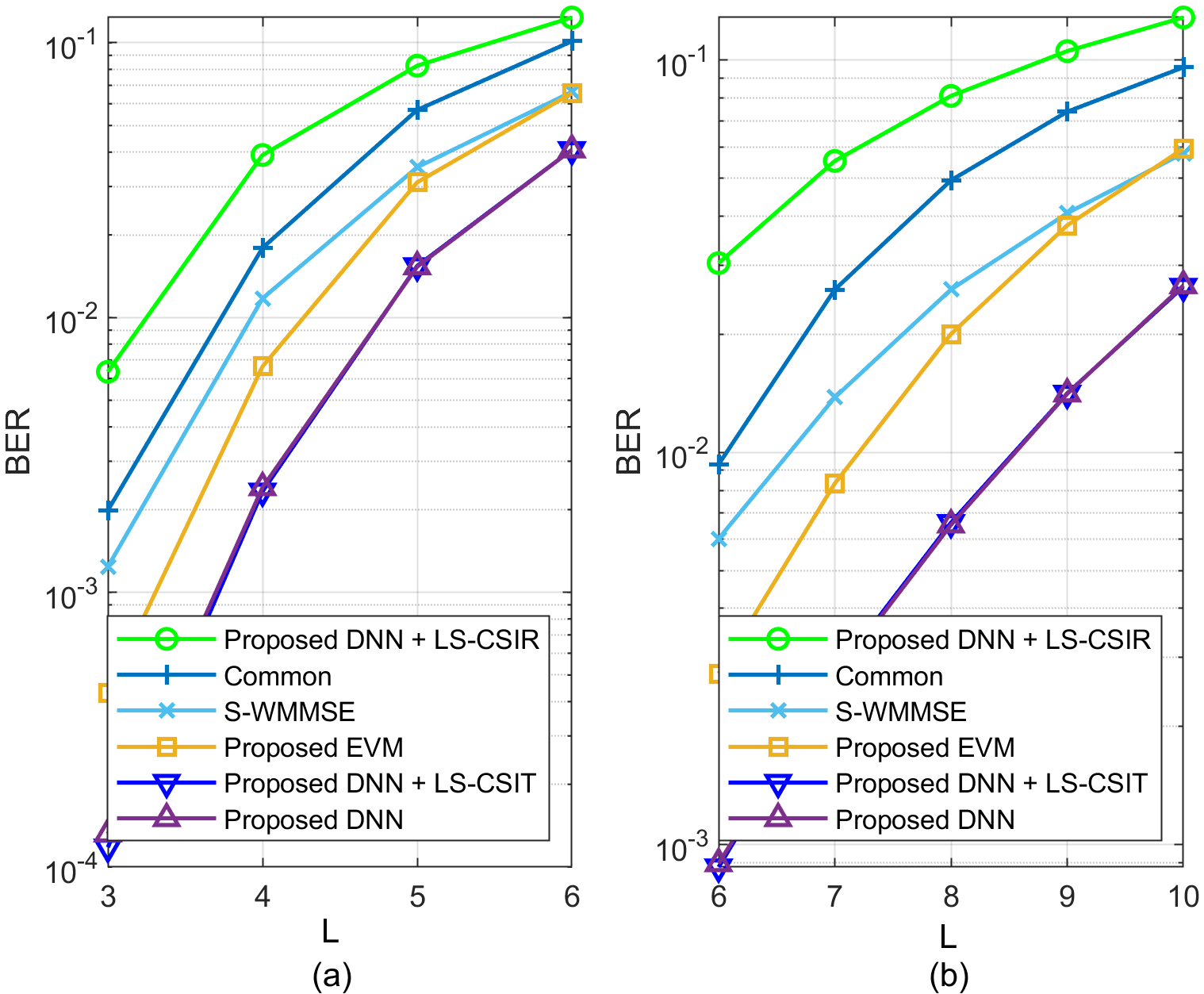}
	\caption{BER vs $L$ using 4096 QAM at ${\rm{SNR}}=38$ dB: (a) setup A; (b) setup B.}\label{fig::res_rf}
\end{figure}

\Figref{fig::res_rf}{} displays the BER performance with varying numbers of streams $L$, with an SNR set at 38 dB to maintain the BER within the range of $10^{-2}$ to $10^{-3}$. Our proposed solution outperforms all baselines in every scenario. We also include two curves where the CSIT or CSIR estimation module in our DNN solution is replaced by the LS estimator. The results show that the end-to-end BER performance is highly sensitive to CSIR accuracy, while the performance remains stable even when the NMSE of CSIT increases by 100 times.

\subsubsection{BER for Different SNRs}
\begin{figure}[!t]
    \centering
	   \subfigure[$N_\mathrm{t}=8, N_\mathrm{r}=8, L=4$]    {
	    	\includegraphics[trim={0 0 0 0},clip,width=0.8\linewidth]{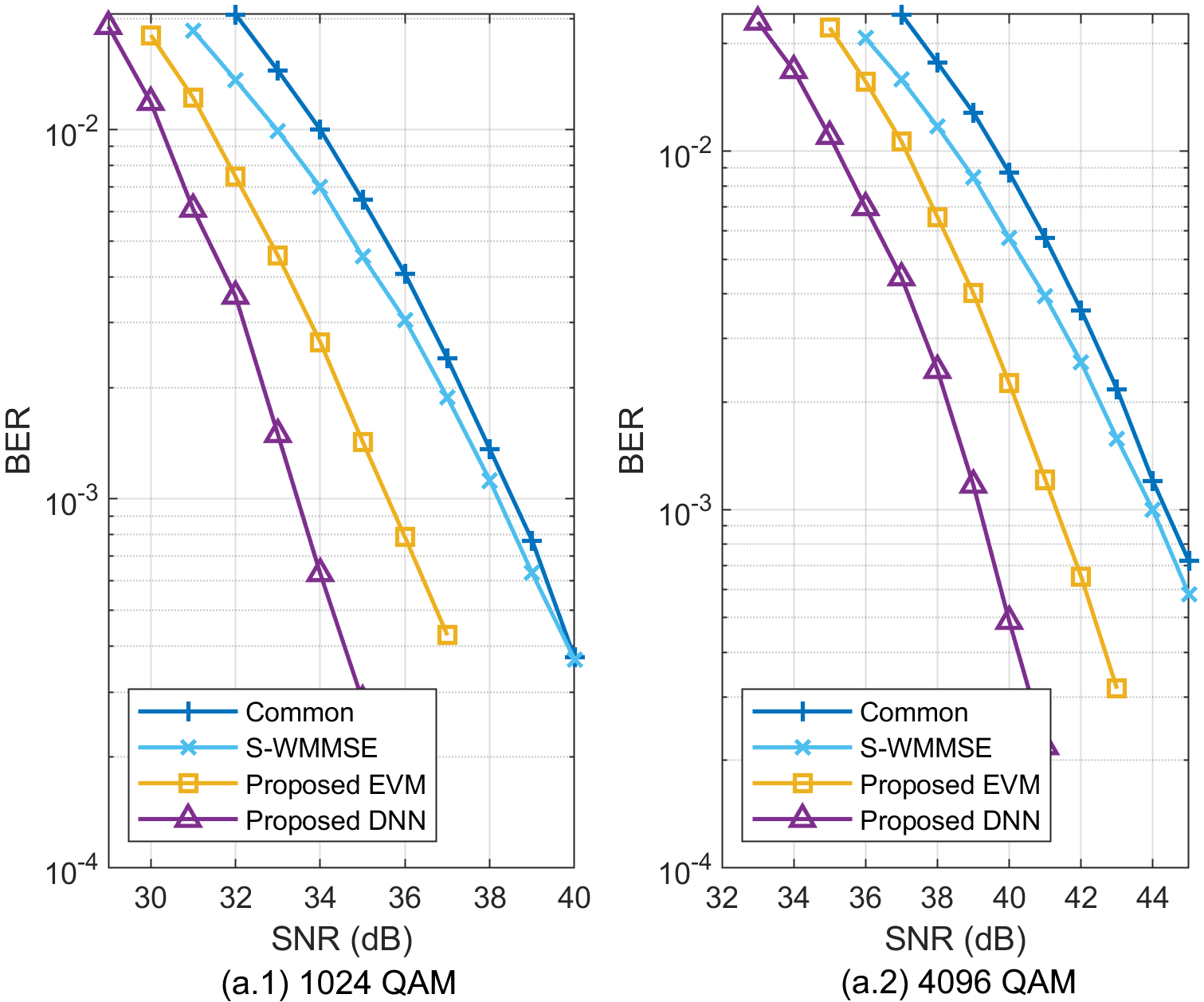}
	    	}
          \subfigure[$N_\mathrm{t}=16, N_\mathrm{r}=16, L=8$]	   {
	    	\includegraphics[trim={0 0 0 0},clip,width=0.8\linewidth]{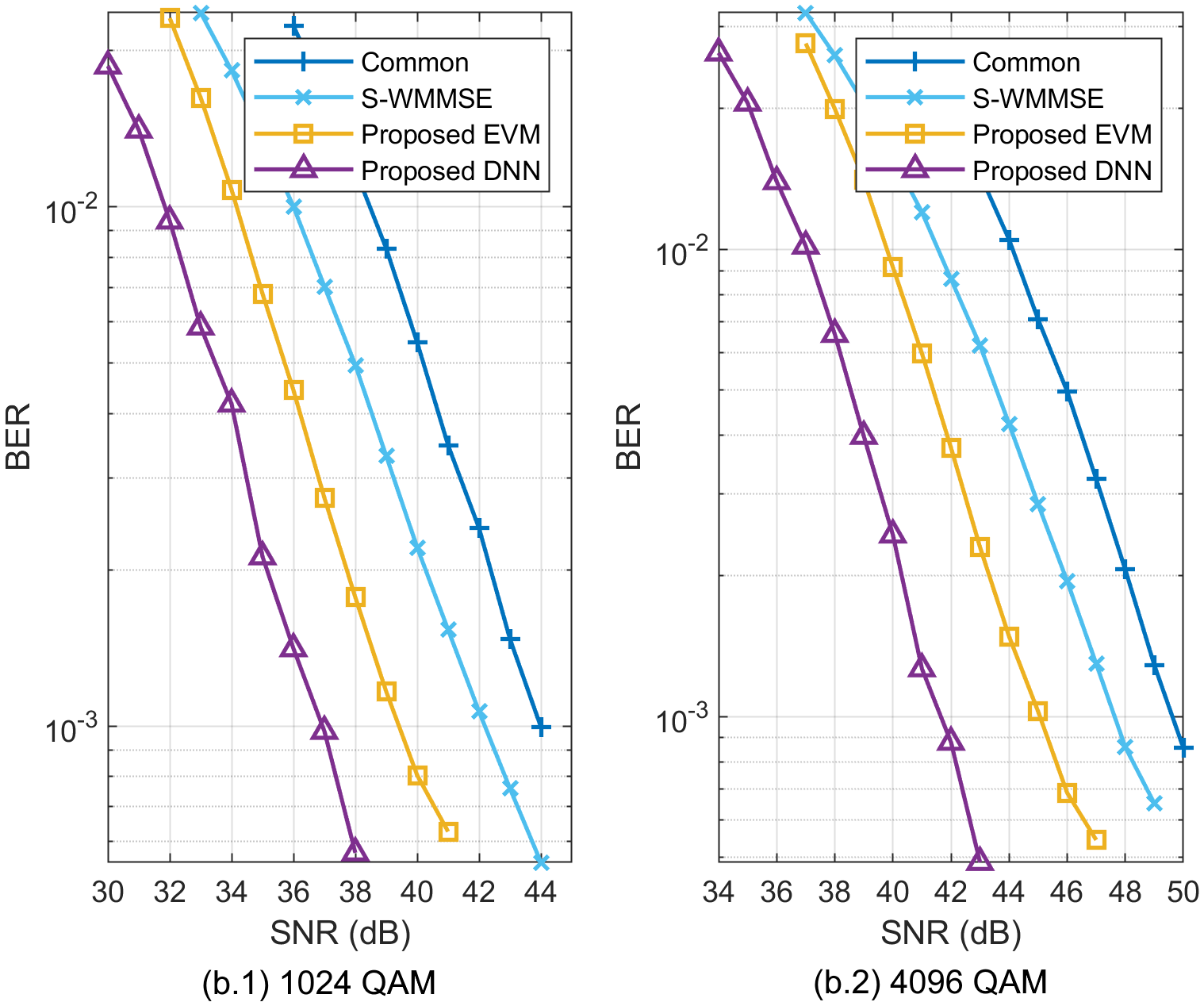}
		}
	\caption{BER vs SNR for proposed and baseline schemes.}\label{fig::res_ber}
\end{figure}

Finally, \Figref{fig::res_ber}{} illustrates the BER performance across varying SNRs, with $L = N_{\rm t}/2$ for both 1024 QAM and 4096 QAM. At a BER threshold of $10^{-3}$, our proposed optimization scheme achieves a 2.5 dB improvement over S-WMMSE by shifting the objective from achievable rate to EVM. Furthermore, our DNN solution offers an additional gain of over 2.5 dB compared to the optimization scheme, resulting in a total improvement of 5 dB over the baselines, thanks to the joint design of the transmitter and receiver.


%

\section{Conclusions}
In this work, we propose a sparse precoding design aimed at reducing pilot overhead for CSIR estimation in MIMO-OFDM systems utilizing higher-order modulation. Our sparse precoder design achieves up to an 8x reduction in pilot overhead by enforcing sparsity in the delay domain. Specifically, we introduce a DNN-based end-to-end solution that includes a propagation channel estimation module, a precoder design module, and an effective channel estimation module. The DNN weights are jointly trained using a model-assisted variational Bayesian training framework, which incorporates domain knowledge to ensure an interpretable design with improved performance. Simulations demonstrate that our solution achieves more than 5 dB improvement over various baseline solutions.

\bibliographystyle{IEEEtran}
\bibliography{IEEEabrv, bibliography}

\begin{thebibliography}{10}
\providecommand{\url}[1]{#1}
\csname url@samestyle\endcsname
\providecommand{\newblock}{\relax}
\providecommand{\bibinfo}[2]{#2}
\providecommand{\BIBentrySTDinterwordspacing}{\spaceskip=0pt\relax}
\providecommand{\BIBentryALTinterwordstretchfactor}{4}
\providecommand{\BIBentryALTinterwordspacing}{\spaceskip=\fontdimen2\font plus
\BIBentryALTinterwordstretchfactor\fontdimen3\font minus
  \fontdimen4\font\relax}
\providecommand{\BIBforeignlanguage}[2]{{%
\expandafter\ifx\csname l@#1\endcsname\relax
\typeout{** WARNING: IEEEtran.bst: No hyphenation pattern has been}%
\typeout{** loaded for the language `#1'. Using the pattern for}%
\typeout{** the default language instead.}%
\else
\language=\csname l@#1\endcsname
\fi
#2}}
\providecommand{\BIBdecl}{\relax}
\BIBdecl

\bibitem{heath2016overview}
R.~W. Heath, N.~Gonzalez-Prelcic, S.~Rangan, W.~Roh, and A.~M. Sayeed, ``An
  overview of signal processing techniques for millimeter wave {{MIMO}}
  systems,'' \emph{IEEE J. Sel. Topics Signal Process.}, vol.~10, no.~3, pp.
  436--453, 2016.

\bibitem{bolcskei2006mimo}
H.~Bolcskei, ``{{MIMO}-{OFDM}} wireless systems: basics, perspectives, and
  challenges,'' \emph{IEEE Wireless Commun.}, vol.~13, no.~4, pp. 31--37, 2006.

\bibitem{harkat2022survey}
H.~Harkat, P.~Monteiro, A.~Gameiro, F.~Guiomar, and H.~Farhana Thariq~Ahmed,
  ``A survey on {{MIMO}-{OFDM}} systems: Review of recent trends,''
  \emph{Signals}, vol.~3, no.~2, pp. 359--395, 2022.

\bibitem{liu2016exploiting}
A.~Liu, V.~K. Lau, and W.~Dai, ``Exploiting burst-sparsity in massive {{MIMO}}
  with partial channel support information,'' \emph{IEEE Trans. Wireless
  Commun.}, vol.~15, no.~11, pp. 7820--7830, 2016.

\bibitem{liu2016joint}
A.~Liu, V.~Lau, and W.~Dai, ``Joint burst {LASSO} for sparse channel estimation
  in multi-user massive {{MIMO}},'' in \emph{Proc. IEEE ICC}, 2016, pp. 1--6.

\bibitem{liu2018downlink}
A.~Liu, L.~Lian, V.~K. Lau, and X.~Yuan, ``Downlink channel estimation in
  multiuser massive {{MIMO}} with hidden {Markovian} sparsity,'' \emph{IEEE
  Trans. Signal Process.}, vol.~66, no.~18, pp. 4796--4810, 2018.

\bibitem{dahlman20205g}
E.~Dahlman, S.~Parkvall, and J.~Skold, \emph{{5G} {NR}: The next generation
  wireless access technology}.\hskip 1em plus 0.5em minus 0.4em\relax Academic
  Press, 2020.

\bibitem{mohammadian2016deterministic}
R.~Mohammadian, A.~Amini, and B.~H. Khalaj, ``Deterministic pilot design for
  sparse channel estimation in {MISO}/multi-user {OFDM} systems,'' \emph{IEEE
  Trans. Wireless Commun.}, vol.~16, no.~1, pp. 129--140, 2016.

\bibitem{li2022adaptive}
T.~Li, N.~Noels, K.~Y. Kapusuz, S.~Lemey, H.~Rogier, and H.~Steendam,
  ``Adaptive pilot allocation for estimating sparse uplink {MU}-{MIMO}-{OFDM}
  channels,'' \emph{IEEE Trans. Wireless Commun.}, vol.~21, no.~10, pp.
  8230--8244, 2022.

\bibitem{qi2011optimized}
C.~Qi and L.~Wu, ``Optimized pilot placement for sparse channel estimation in
  {OFDM} systems,'' \emph{IEEE Signal Process. Lett.}, vol.~18, no.~12, pp.
  749--752, 2011.

\bibitem{chen2013efficient}
J.-C. Chen, C.-K. Wen, and P.~Ting, ``An efficient pilot design scheme for
  sparse channel estimation in {OFDM} systems,'' \emph{IEEE Commun. Lett.},
  vol.~17, no.~7, pp. 1352--1355, 2013.

\bibitem{wang2022channel}
Y.~Wang, C.~Qi, P.~Li, Z.~Lu, and P.~Lu, ``Channel estimation for wideband
  {mmWave {MIMO} {OFDM}} system exploiting block sparsity,'' \emph{IEEE Commun.
  Lett.}, vol.~26, no.~4, pp. 897--901, 2022.

\bibitem{srivastava2021bayesian}
S.~Srivastava, R.~K. Singh, A.~K. Jagannatham, and L.~Hanzo, ``Bayesian
  learning aided simultaneous row and group sparse channel estimation in
  orthogonal time frequency space modulated {MIMO} systems,'' \emph{IEEE Trans.
  Commun.}, vol.~70, no.~1, pp. 635--648, 2021.

\bibitem{zhu2019channel}
F.~Zhu, A.~Liu, and V.~K. Lau, ``Channel estimation and localization for mmwave
  systems: A sparse bayesian learning approach,'' in \emph{Proc. IEEE ICC)},
  2019, pp. 1--6.

\bibitem{chen2018massive}
L.~Chen and X.~Yuan, ``Massive {MIMO}-{OFDM} channel estimation via structured
  turbo compressed sensing,'' in \emph{IEEE ICC}, 2018, pp. 1--6.

\bibitem{christensen2008weighted}
S.~S. Christensen, R.~Agarwal, E.~De~Carvalho, and J.~M. Cioffi, ``Weighted
  sum-rate maximization using weighted {MMSE} for {{MIMO}-BC} beamforming
  design,'' \emph{IEEE Trans. Wireless Commun.}, vol.~7, no.~12, pp.
  4792--4799, 2008.

\bibitem{shi2011iteratively}
Q.~Shi, M.~Razaviyayn, Z.-Q. Luo, and C.~He, ``An iteratively weighted {{MMSE}}
  approach to distributed sum-utility maximization for a {{MIMO}} interfering
  broadcast channel,'' \emph{IEEE Trans. Signal Process.}, vol.~59, no.~9, pp.
  4331--4340, 2011.

\bibitem{perahia2013next}
E.~Perahia and R.~Stacey, \emph{Next generation wireless {LANs}: 802.11 n and
  802.11 ac}.\hskip 1em plus 0.5em minus 0.4em\relax Cambridge university
  press, 2013.

\bibitem{ali2017beamforming}
E.~Ali, M.~Ismail, R.~Nordin, and N.~F. Abdulah, ``Beamforming techniques for
  massive {{MIMO}} systems in {{5G}}: overview, classification, and trends for
  future research,'' \emph{Frontiers of Inf. Technol. \& Electron. Eng.},
  vol.~18, pp. 753--772, 2017.

\bibitem{tiwari2023advancing}
P.~Tiwari, V.~Gahlaut, M.~Kaushik, P.~Rani, A.~Shastri, and B.~Singh,
  ``Advancing {5G} connectivity: a comprehensive review of {{MIMO}} antennas
  for {5G} applications,'' \emph{International J. of Antennas Propag.}, vol.
  2023, no.~1, p. 5906721, 2023.

\bibitem{ravindran2008limited}
N.~Ravindran and N.~Jindal, ``Limited feedback-based block diagonalization for
  the {MIMO} broadcast channel,'' \emph{IEEE J. Sel. Areas Commun.}, vol.~26,
  no.~8, pp. 1473--1482, 2008.

\bibitem{sung2009generalized}
H.~Sung, S.-R. Lee, and I.~Lee, ``Generalized channel inversion methods for
  multiuser {MIMO} systems,'' \emph{IEEE Trans. Commun.}, vol.~57, no.~11, pp.
  3489--3499, 2009.

\bibitem{joham2005linear}
M.~Joham, W.~Utschick, and J.~A. Nossek, ``Linear transmit processing in {MIMO}
  communication systems,'' \emph{IEEE Trans. Signal Process.}, vol.~53, no.~8,
  pp. 2700--2712, 2005.

\bibitem{moon2014channel}
S.-H. Moon, S.-R. Lee, J.-S. Kim, and I.~Lee, ``Channel quantization for block
  diagonalization with limited feedback in multiuser {MIMO} downlink
  channels,'' \emph{J. of Commun. and Networks}, vol.~16, no.~1, pp. 1--9,
  2014.

\bibitem{zhang2009robust}
C.~Zhang, W.~Xu, and M.~Chen, ``Robust {MMSE} beamforming for multiuser {MISO}
  systems with limited feedback,'' \emph{IEEE Signal Process. Lett.}, vol.~16,
  no.~7, pp. 588--591, 2009.

\bibitem{vucic2009robust}
N.~Vucic, H.~Boche, and S.~Shi, ``Robust transceiver optimization in downlink
  multiuser {MIMO} systems,'' \emph{IEEE Trans. Signal Process.}, vol.~57,
  no.~9, pp. 3576--3587, 2009.

\bibitem{dabbagh2008multiple}
A.~D. Dabbagh and D.~J. Love, ``Multiple antenna {MMSE} based downlink
  precoding with quantized feedback or channel mismatch,'' \emph{IEEE Trans.
  Commun.}, vol.~56, no.~11, pp. 1859--1868, 2008.

\bibitem{fritzsche2013robust}
R.~Fritzsche and G.~P. Fettweis, ``Robust sum rate maximization in the
  multi-cell {MU}-{MIMO} downlink,'' in \emph{Proc. IEEE WCNC}, 2013, pp.
  3180--3184.

\bibitem{precodedelay2019WCL}
V.~Ramireddy, M.~Grossmann, M.~Landmann, and G.~D. Galdo, ``Sub-band versus
  space-delay precoding for wideband {mmWave} channels,'' \emph{IEEE Wireless
  Commun. Lett.}, vol.~8, no.~1, pp. 193--196, 2019.

\bibitem{zhang2023cross}
Y.~Zhang, A.-A. Lu, B.~Liu, X.~Gao, and X.-G. Xia, ``Cross-subcarrier precoder
  design for massive {MIMO}-{OFDM} downlink,'' in \emph{Proc. IEEE
  VTC2023-Fall}, 2023, pp. 1--6.

\bibitem{sohrabi2021deep}
F.~Sohrabi, K.~M. Attiah, and W.~Yu, ``Deep learning for distributed channel
  feedback and multiuser precoding in {FDD} massive {{MIMO}},'' \emph{IEEE
  Trans. Wireless Commun.}, vol.~20, no.~7, pp. 4044--4057, 2021.

\bibitem{jang2019deep}
J.~Jang, H.~Lee, S.~Hwang, H.~Ren, and I.~Lee, ``Deep learning-based limited
  feedback designs for {{MIMO}} systems,'' \emph{IEEE Wireless Commun. Lett.},
  vol.~9, no.~4, pp. 558--561, 2019.

\bibitem{wu2022deep}
M.~Wu, Z.~Gao, Z.~Gao, D.~Wu, Y.~Yang, and Y.~Huang, ``Deep learning-based
  hybrid precoding for {FDD} massive {{MIMO}-{OFDM}} systems with a limited
  pilot and feedback overhead,'' in \emph{Proc. IEEE ICC Workshops}.\hskip 1em
  plus 0.5em minus 0.4em\relax IEEE, 2022, pp. 318--323.

\bibitem{zhang2022deep}
M.~Zhang, J.~Gao, and C.~Zhong, ``A deep learning-based framework for low
  complexity multiuser {{MIMO}} precoding design,'' \emph{IEEE Trans. Wireless
  Commun.}, vol.~21, no.~12, pp. 11\,193--11\,206, 2022.

\bibitem{attiah2022deep}
K.~M. Attiah, F.~Sohrabi, and W.~Yu, ``Deep learning for channel sensing and
  hybrid precoding in {{TDD}} massive {{MIMO} {OFDM}} systems,'' \emph{IEEE
  Trans. Wireless Commun.}, vol.~21, no.~12, pp. 10\,839--10\,853, 2022.

\bibitem{pellaco2020deep}
L.~Pellaco, M.~Bengtsson, and J.~Jald{\'e}n, ``Deep unfolding of the weighted
  {MMSE} beamforming algorithm,'' \emph{arXiv preprint arXiv:2006.08448}, 2020.

\bibitem{jang2022deep}
J.~Jang, H.~Lee, I.-M. Kim, and I.~Lee, ``Deep learning for multi-user {{MIMO}}
  systems: Joint design of pilot, limited feedback, and precoding,'' \emph{IEEE
  Trans. Commun.}, vol.~70, no.~11, pp. 7279--7293, 2022.

\bibitem{shi2023robust}
J.~Shi, A.-A. Lu, W.~Zhong, X.~Gao, and G.~Y. Li, ``Robust {W{MMSE}} precoder
  with deep learning design for massive {{MIMO}},'' \emph{IEEE Trans. Commun.},
  vol.~71, no.~7, pp. 3963--3976, 2023.

\bibitem{jin2023model}
W.~Jin, J.~Zhang, C.-K. Wen, and S.~Jin, ``Model-driven deep learning for
  hybrid precoding in millimeter wave {{MU}-{MIMO}} system,'' \emph{IEEE Trans.
  Commun.}, 2023.

\bibitem{hu2021two}
Q.~Hu, Y.~Cai, K.~Kang, G.~Yu, J.~Hoydis, and Y.~C. Eldar, ``Two-timescale
  end-to-end learning for channel acquisition and hybrid precoding,''
  \emph{IEEE J. Sel. Areas Commun.}, vol.~40, no.~1, pp. 163--181, 2021.

\bibitem{monga2021algorithm}
V.~Monga, Y.~Li, and Y.~C. Eldar, ``Algorithm unrolling: Interpretable,
  efficient deep learning for signal and image processing,'' \emph{IEEE Signal
  Process. Mag.}, vol.~38, no.~2, pp. 18--44, 2021.

\bibitem{forenza2007simplified}
A.~Forenza, D.~J. Love, and R.~W. Heath, ``Simplified spatial correlation
  models for clustered {MIMO} channels with different array configurations,''
  \emph{IEEE Trans. Veh. Technol.}, vol.~56, no.~4, pp. 1924--1934, 2007.

\bibitem{hoydis2022sionna}
J.~Hoydis, S.~Cammerer, F.~A. Aoudia, A.~Vem, N.~Binder, G.~Marcus, and
  A.~Keller, ``Sionna: An open-source library for next-generation physical
  layer research,'' \emph{arXiv preprint arXiv:2203.11854}, 2022.

\bibitem{guo2023robust}
H.~Guo and V.~K. Lau, ``Robust deep learning for uplink channel estimation in
  cellular network under inter-cell interference,'' \emph{IEEE J. Sel. Areas
  Commun.}, vol.~41, no.~6, pp. 1873--1887, 2023.

\bibitem{kingma2013auto}
D.~P. Kingma and M.~Welling, ``Auto-encoding variational bayes,'' \emph{arXiv
  preprint arXiv:1312.6114}, 2013.

\bibitem{jang2016categorical}
E.~Jang, S.~Gu, and B.~Poole, ``Categorical reparameterization with
  gumbel-softmax,'' \emph{arXiv preprint arXiv:1611.01144}, 2016.

\bibitem{tensorflow2015}
\BIBentryALTinterwordspacing
M.~Abadi, A.~Agarwal \emph{et~al.}, ``{TensorFlow}: Large-scale machine
  learning on heterogeneous systems,'' 2015, software available from
  tensorflow.org. [Online]. Available: \url{https://www.tensorflow.org/}
\BIBentrySTDinterwordspacing

\bibitem{kingma2014adam}
D.~P. Kingma and J.~Ba, ``Adam: A method for stochastic optimization,''
  \emph{arXiv preprint arXiv:1412.6980}, 2014.

\bibitem{tropp2007signal}
J.~A. Tropp and A.~C. Gilbert, ``Signal recovery from random measurements via
  orthogonal matching pursuit,'' \emph{IEEE Trans. Inf. Theory}, vol.~53,
  no.~12, pp. 4655--4666, 2007.

\bibitem{ranstam2018lasso}
J.~Ranstam and J.~A. Cook, ``{LASSO} regression,'' \emph{Journal of British
  Surgery}, vol. 105, no.~10, pp. 1348--1348, 2018.

\end{thebibliography}

\end{document}